\documentclass{article}
\usepackage{iclr2027_conference,times}
\iclrfinalcopy
\usepackage[utf8]{inputenc}
\usepackage[T1]{fontenc}
\usepackage{amsmath,amssymb,booktabs,graphicx,xcolor}
\usepackage{hyperref,url,wrapfig}
\hypersetup{hidelinks}
\newcommand{\secondbest}[1]{{\setlength{\fboxsep}{0.5pt}\setlength{\fboxrule}{0.3pt}\fbox{#1}}}
\title{A High-Compression Transformer VAE for Efficient and Controllable 3D Medical Image Generation with Diffusion Transformers}
\author{Bo Liu$^{1}$, Bingliang Zhang$^{2}$, Yubo Wang$^{1}$, Yunxiang Li$^{1}$, Yang Yang$^{1}$,\\ \bfseries Janine M. Lupo$^{1}$, Hui Lin$^{1}$\\
{\normalfont $^{1}$University of California, San Francisco\quad $^{2}$California Institute of Technology}}

\begin{document}
\raggedbottom
\maketitle
\lhead{Preprint}
\suppressfloats[t]

\begin{abstract}
Many medical image processing tasks, such as completing missing regions or correcting acquisition artifacts, can be cast as volumetric generation conditioned on observations of the same patient. Diffusion Transformers offer scalable capacity and flexible conditioning for this setting, but they operate on a learned image representation that must preserve fine anatomical and pathological detail while keeping complete three-dimensional volumes tractable. We bring Transformer variational autoencoding and multimodal diffusion together to address this challenge. Trained on BrainTumor-256K, a 249M Transformer VAE encodes and decodes $256^3$ volumes 7.6$\times$ faster than a comparably sized CNN trained on the same collection. Their PSNR difference is near zero with training-sized crop inputs, but the Transformer leads by 0.749 dB with full-volume inputs and an unchanged scoring region. Increasing Transformer capacity to 499M improves PSNR across all four brain MRI evaluation sets while retaining 3.8$\times$ lower encoding–decoding latency than the CNN. Fast encoding also allows conditions to be constructed dynamically in image space during latent diffusion training. Within the compressed representation, a shared Diffusion Transformer integrates text and spatial references for volumetric synthesis, completion and restoration. The brain-trained autoencoder further transfers to non-brain MRI and CT without target-domain fine-tuning, extending its use beyond the anatomy and modality on which it was learned. The framework provides a reusable representation and conditioning interface for efficient, controllable medical image generation. Code is available at \url{https://github.com/OncoMMAI/brainmri-vit-vae3d}.

\end{abstract}

\begin{figure}[t]
\centering
\includegraphics[width=\linewidth]{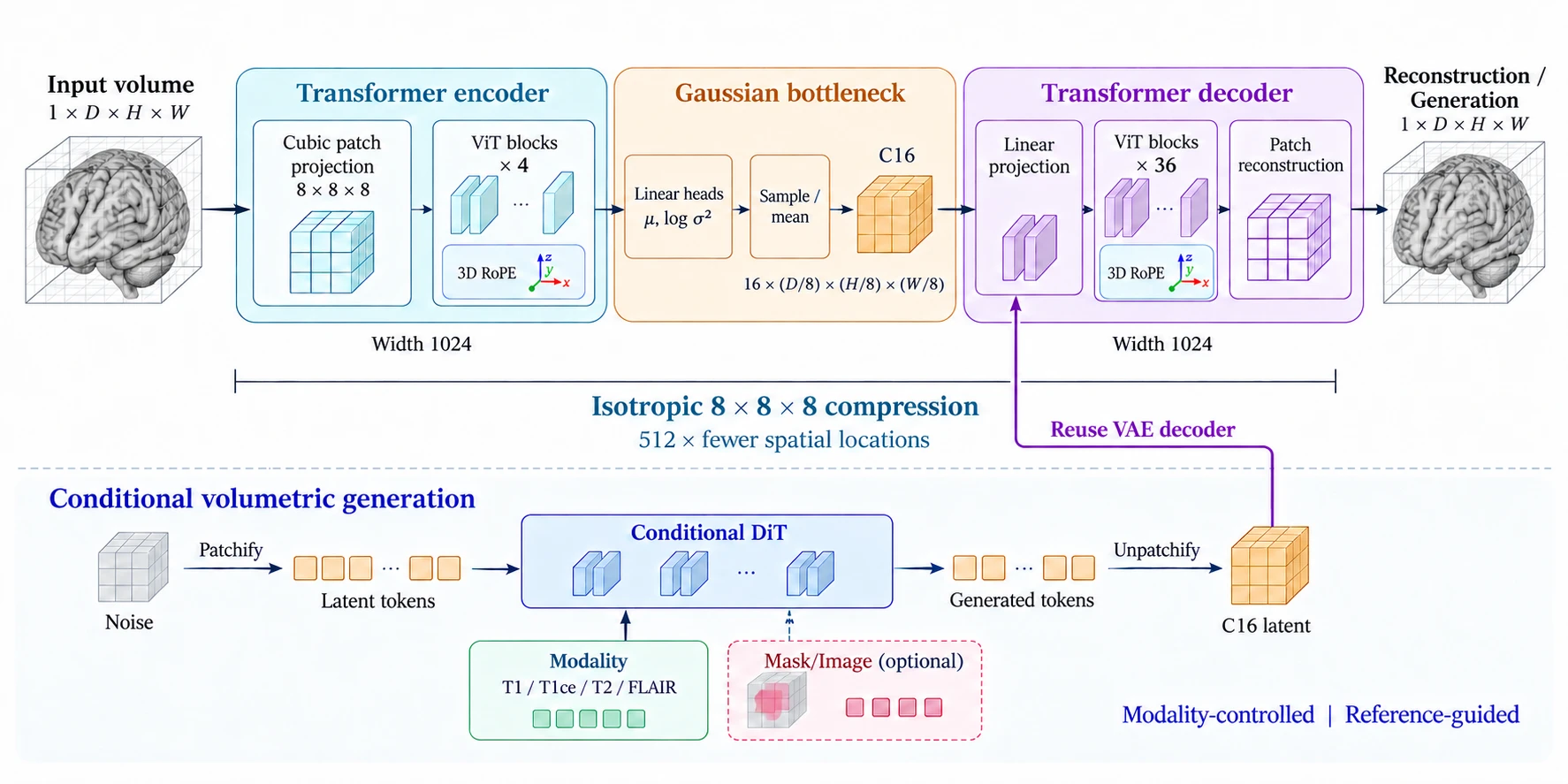}
\caption{Volumetric Transformer VAE with isotropic $8\times8\times8$ compression and 3D RoPE. A conditional DiT generates latent tokens from noise using modality conditions and optional mask or image references, then reuses the VAE decoder. Brain renderings are schematic.}
\label{fig:architecture}
\end{figure}

\section{Introduction}
Multimodal Diffusion Transformers condition generation on reference images as well as text, allowing a single model to perform many image-to-image tasks \citep{flux2024,qwenimage2025}. In medical imaging, the same formulation covers problems that are usually addressed by separate models, such as completing missing regions, correcting acquisition artifacts and synthesizing unavailable contrasts. Here the reference is a partial or degraded observation of the same patient, and the generated volume must remain consistent with it while preserving fine anatomical and pathological detail. Latent diffusion makes such generation tractable by moving computation into a compressed representation \citep{rombach2022ldm}, and Diffusion Transformers provide a common token interface for text and visual conditions \citep{peebles2023dit,esser2024rectified}. In three dimensions, however, the token count grows with volume and the cost of self-attention grows quadratically with the token count; each spatial reference further adds a volumetric sequence of the same length. The degree of spatial compression in the latent grid is therefore central to making controllable volumetric diffusion practical.

The cost of the autoencoder also enters diffusion training when conditions are constructed in image space. For tasks such as inpainting or motion correction, each masked or degraded volume must be encoded before it is passed to the DiT. Encoding these conditions online adds computational overhead; a fast encoder reduces this cost and makes resampling masks and degradations at every training step more practical. Existing medical autoencoders support MRI synthesis and controllable CT generation \citep{pinaya2022brain,khader2022medical,guo2024maisi,wang2024meddiffusion}, but they typically adopt milder compression or convolutional backbones whose cost scales with the number of voxels, which makes high-capacity full-volume processing expensive.

Building on scalable visual tokenization \citep{chen2024dcae,vitok2025,vitokv22026}, we develop a Transformer VAE that patchifies the input by $8\times8\times8$ at its first layer, so that all subsequent computation operates on the latent grid ($32^3$ positions for a $256^3$ volume). Equal-axis compression preserves the isotropy of the input grid, and 3D rotary position encoding supports inference beyond the $128^3$ training crop, with inputs padded to multiples of eight. We train the model on BrainTumor-256K, a collection of 256K brain MRI volumes. Compared with a CNN VAE of matched size, training data and latent interface, the Transformer achieves comparable fidelity on training-sized crops while encoding and decoding $7.6\times$ faster at $256^3$. The Transformer supports full-volume processing at substantially lower measured latency, and increasing its decoder capacity improves reconstruction while retaining a latency advantage over the trained CNN.

Using this representation, we train a multimodal DiT that conditions on text, lesion masks and observed images through a shared reference pathway (Figure~\ref{fig:architecture}). Fast encoding allows masks and degradations to be generated online during training, enabling a single model to perform six tasks ranging from sequence-controlled synthesis to motion correction.

Our contributions are:
\begin{enumerate}
\setlength{\itemsep}{2pt}\setlength{\parsep}{0pt}
\item \textbf{An efficient high-compression volumetric VAE.} Running the Transformer on a latent grid downsampled eightfold along each spatial axis makes encoding and decoding several times faster than a comparably sized CNN, and a larger decoder improves reconstruction while retaining this latency advantage.
\item \textbf{A faithful and reusable medical representation.} Trained on brain MRI, the 500M model achieves the highest fidelity among the evaluated autoencoders on four brain test sets and transfers to other MRI anatomies and to CT without fine-tuning.
\item \textbf{One conditional model for multiple medical imaging tasks.} A 2B multimodal DiT performs synthesis, completion and restoration using text conditions and, where applicable, spatial references constructed online during training.
\end{enumerate}

\section{Related Work}
\paragraph{High-compression image representations.}
The usefulness of a compressed representation depends on both what it preserves and how easily a generative prior can model it. Variational and vector-quantized autoencoders establish continuous and discrete interfaces \citep{kingma2013vae,oord2017vqvae}, with perceptual and adversarial objectives improving image recovery under compression \citep{esser2021taming}. Latent diffusion and Diffusion Transformers make this interface a determinant of generative cost \citep{rombach2022ldm,peebles2023dit}. DC-AE, ViTok and ViTok-v2 explore deeper compression, model scaling and native-resolution Transformer autoencoding \citep{chen2024dcae,vitok2025,vitokv22026}, while representation and prior alignment address latent properties relevant to generation \citep{zheng2025rae,prioraligned2026}. Extending these ideas to medical volumes requires preserving spatial anatomy while keeping representation learning and downstream diffusion tractable.

\paragraph{Volumetric medical autoencoders.}
Volumetric latent models already support brain MRI generation and controllable CT synthesis \citep{pinaya2022brain,khader2022medical,guo2024maisi}. Their autoencoders embody different choices about spatial compression, patch processing and anatomy: 3D MedDiffusion uses a patch-volume representation, MedVAE studies medical autoencoding, and NeuroQuant develops modality-aware anatomical VQ representations \citep{wang2024meddiffusion,medvae2025,anatomical2026}. CONFLUX and whole-volume translation extend latent modeling to chest CT synthesis and volumetric translation \citep{conflux2026,wholevolume2026}. Foundation VAEs for CT pursue reuse of frozen image/video autoencoders for reconstruction, augmentation and generation \citep{foundationct2026}. Our focus is the connection between medically trained representation capacity, computational efficiency and a shared conditional interface for volumetric generation.

\paragraph{Video autoencoders for volumetric data.}
CogVideoX, HunyuanVideo and Wan provide pretrained video VAEs that can be adapted to medical volumes \citep{yang2024cogvideox,kong2024hunyuan,wan2025}. Their temporal--spatial designs contrast with our equal-axis compression and bidirectional processing. Cosmos also provides nominal $8\times8\times8$ compression with C16 latents, using an AE formulation without a KL prior and temporally causal processing \citep{nvidia2025cosmos}. We include released video VAEs as reconstruction and latency baselines.

\paragraph{Image-conditioned generation.}
Multimodal rectified-flow Transformers provide a shared architecture for text and visual conditioning \citep{esser2024rectified}. FLUX combines multimodal and parallel Transformer blocks, while Qwen-Image integrates text and reference-image conditioning \citep{flux2024,qwenimage2025}. We adopt this general conditioning paradigm for volumetric MRI, using a compact VAE grid to make three-dimensional target and reference representations tractable.

\begin{table}[!tb]
\caption{Brain MRI autoencoding fidelity (mean PSNR in dB $\uparrow$) and latent compression. Rows are ordered by BrainTumor-256K PSNR; bold and boxed values denote best and second-best scores.}
\label{tab:brain}
\centering\small\setlength{\tabcolsep}{3pt}\resizebox{\linewidth}{!}{\begin{tabular}{lrrrrcrrr}
\toprule
Model & BraTS & OASIS & IXI & \shortstack{BrainTumor-\\256K} & Compression & \shortstack{Spatial\\Reduction} & Channel & \shortstack{Element\\Reduction} \\
\midrule
ViT-VAE 500M & \textbf{33.53} & \textbf{32.16} & \textbf{35.46} & \textbf{33.80} & $8\times8\times8$ & 512$\times$ & 16 & 32$\times$ \\
ViT-VAE 250M & \secondbest{32.55} & \secondbest{31.13} & \secondbest{34.08} & \secondbest{32.95} & $8\times8\times8$ & 512$\times$ & 16 & 32$\times$ \\
CNN VAE 247M & 32.46 & 30.06 & 33.82 & 32.18 & $8\times8\times8$ & 512$\times$ & 16 & 32$\times$ \\
CogVideoX & 29.62 & 28.71 & 24.46 & 28.91 & $4\times8\times8$ & 256$\times$ & 16 & 16$\times$ \\
HunyuanVideo & 29.83 & 28.72 & 30.79 & 28.78 & $4\times8\times8$ & 256$\times$ & 16 & 16$\times$ \\
Wan & 29.18 & 28.20 & 30.30 & 28.04 & $4\times8\times8$ & 256$\times$ & 16 & 16$\times$ \\
MAISI-MR & 27.65 & 26.12 & 27.75 & 26.81 & $4\times4\times4$ & 64$\times$ & 4 & 16$\times$ \\
MedVAE-f4c1 & 27.04 & 25.67 & 27.85 & 26.16 & $4\times4\times4$ & 64$\times$ & 1 & 64$\times$ \\
MAISI-CT & 26.59 & 25.39 & 25.54 & 26.16 & $4\times4\times4$ & 64$\times$ & 4 & 16$\times$ \\
Patch-Volume & 22.80 & 22.01 & 22.80 & 22.03 & $8\times8\times8$ & 512$\times$ & 8 & 64$\times$ \\
MedVAE-f8c1 & 22.28 & 20.23 & 21.88 & 21.91 & $8\times8\times8$ & 512$\times$ & 1 & 512$\times$ \\
Brain-LDM & 19.15 & 17.88 & 18.20 & 21.10 & $8\times8\times8$ & 512$\times$ & 3 & 170.7$\times$ \\
CONFLUX & 21.24 & 14.98 & 24.10 & 19.93 & $8\times8\times8$ & 512$\times$ & 16 & 32$\times$ \\
\bottomrule
\end{tabular}
}
\end{table}

\section{High-Compression Autoencoding and Conditional Volumetric Diffusion}
The framework couples a Gaussian autoencoder with a multimodal latent generator. The autoencoder supplies a compact spatial representation and a shared image encoder for conditioning; the DiT learns to synthesize target volumes under text and spatial constraints. 
\subsection{High-compression volumetric representation}
For a grayscale volume $x\in\mathbb{R}^{1\times D\times H\times W}$ with dimensions divisible by eight, the encoder produces a diagonal Gaussian posterior
\begin{equation}
q_\phi(z\mid x)=\mathcal{N}\!\left(\mu_\phi(x),\operatorname{diag}(\sigma_\phi^2(x))\right),\qquad
z\in\mathbb{R}^{16\times D/8\times H/8\times W/8}.
\end{equation}
Training samples $z=\mu+\sigma\odot\epsilon$, where $\epsilon\sim\mathcal{N}(0,I)$; reconstruction evaluation uses the posterior mean. A 3D convolution with kernel and stride $(8,8,8)$ projects voxel patches into spatial tokens, following the patch-embedding approach of ViT \citep{dosovitskiy2021vit}. Global transformer blocks process this compact grid, and the decoder projects the decoded tokens back to voxel patches (Figure~\ref{fig:architecture}).

We distinguish the spatial reduction from the scalar-element reduction:
\begin{equation}
R_{\mathrm{grid}}=8^3=512,\qquad
R_{\mathrm{elements}}=\frac{1\cdot DHW}{16\cdot(D/8)(H/8)(W/8)}=32.
\end{equation}
Appendix~\ref{app:compression} compares latent tensor sizes across interfaces.

Equal downsampling maps input spacing $(s_D,s_H,s_W)$ to latent spacing $(8s_D,8s_H,8s_W)$, so isotropic inputs retain an isotropic latent grid.

At equal C16 bottlenecks, $(8,8,8)$ compression produces half as many latent positions as $(4,8,8)$ compression for divisible input dimensions. At a fixed downstream DiT patch size, this halves the token count and quarters the number of dense-attention interactions. Appendix~\ref{app:compression} gives the accounting, and Section~\ref{sec:dit-resources} examines the resulting training cost.

\subsection{Transformer backbone and variable-shape inference}
The backbone uses a shallow encoder and a deeper decoder, with width 1024, sixteen attention heads and parallel attention \citep{chowdhery2023palm} and SwiGLU processing \citep{shazeer2020glu}. The 249.2M- and 499.3M-parameter variants use four encoder blocks and sixteen or thirty-six decoder blocks, respectively; tables and figures refer to them as ViT-VAE 250M and ViT-VAE 500M. The larger model is used for the primary reconstruction and resource evaluations and serves as the frozen autoencoder for all conditional generation tasks, which share one trained 2B DiT checkpoint. Three-dimensional rotary position encoding applies query--key rotations along the spatial axes \citep{su2021rope,heo2024ropevit}, using coordinates derived from the current token grid.

Training uses $128^3$ crops. At inference, volumes are padded to multiples of eight and cropped back after decoding. Spatial coordinates follow the input grid, allowing the model to process dimensions and aspect ratios beyond the training crop without resizing.

\begin{table}[!tb]
\caption{Encoding--decoding latency (batch one, median ms) and theoretical forward FLOPs. All three models use whole-volume inference; OOM indicates insufficient memory. Full baseline results are in Appendix~\ref{app:implementation}.}
\label{tab:speed}
\centering\small\setlength{\tabcolsep}{5pt}\begin{tabular}{lrrrrr}
\toprule
Model & Params (M) & $128^3$ ms & $256^3$ ms & $384^3$ ms & $256^3$ TFLOPs \\
\midrule
ViT-VAE 500M & 499.3 & 38.7 & 574.0 & 4960.3 & 208.64 \\
ViT-VAE 250M & 249.2 & 19.6 & 288.5 & 2485.5 & 104.29 \\
CNN VAE 247M & 246.5 & 249.7 & 2195.7 & OOM & 387.49 \\
\bottomrule
\end{tabular}

\end{table}

\subsection{Training objective}
We train the VAE with voxelwise, structural and perceptual supervision together with Gaussian prior regularization:
\begin{equation}
\mathcal L=\lambda_{\mathrm{vox}}\operatorname{mean}(x-\hat x)^2
+\lambda_{\mathrm{str}}\bigl(1-\operatorname{SSIM}(x,\hat x)\bigr)
+\lambda_{\mathrm{perc}}\mathcal L_{\mathrm{perc}}(x,\hat x)
+\beta\overline D_{\mathrm{KL}}.
\end{equation}
Here $\overline D_{\mathrm{KL}}$ is averaged over latent elements, and $\mathcal L_{\mathrm{perc}}$ is a slice-based 2D LPIPS loss. Optimization settings are listed in Appendix~\ref{app:training}.

\begin{figure}[!tbp]
\centering
\raggedright\textbf{(a)}\par
\centering\includegraphics[width=\linewidth]{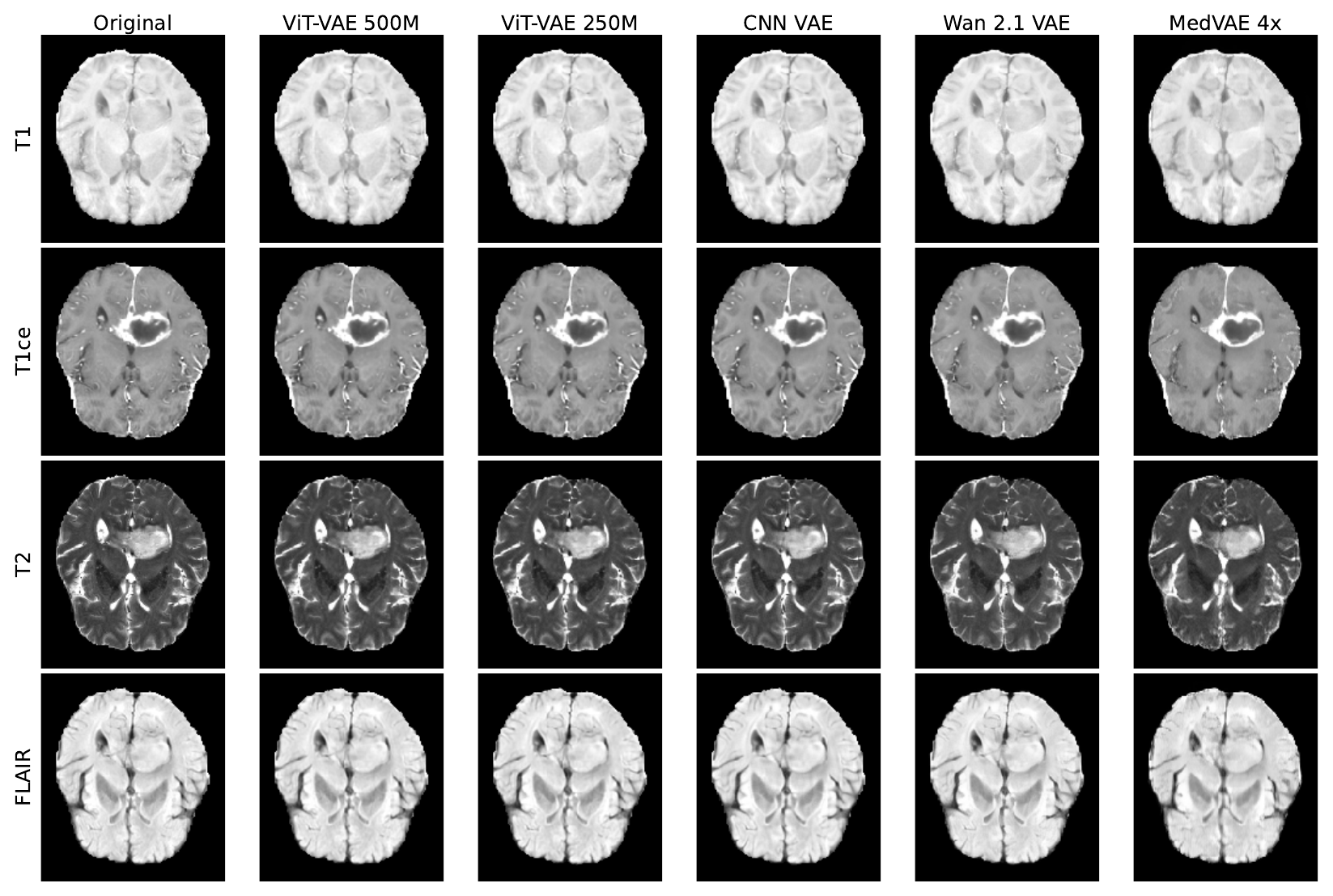}
\par\smallskip\raggedright\textbf{(b)}\par
\centering\includegraphics[width=\linewidth]{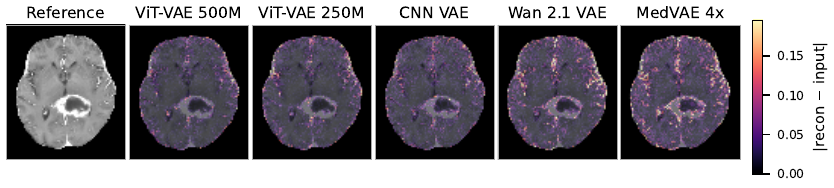}
\caption{Brain MRI autoencoding. \textbf{(a)} Four sequences with shared axial views, crops and intensity limits. \textbf{(b)} T1ce absolute error on a second case, restricted to the brain and shown on a shared color scale. Further examples are in Appendix~\ref{app:examples}.}
\label{fig:qualitative}
\end{figure}

\subsection{Conditional diffusion in the volumetric latent space}
\label{sec:conditional-method}
In this representation, a multimodal Diffusion Transformer learns conditional volumetric synthesis through a rectified-flow formulation \citep{esser2024rectified}. Target latents are patchified into visual tokens, while text and modality embeddings specify the requested sequence and task. Spatial references are encoded on an aligned grid, with multiple references fused at each spatial location. The fused reference tokens are concatenated with the target tokens, doubling the visual sequence length relative to the target alone. Compression thus limits the attention cost of both the generated volume and its spatial conditions. Visual and conditioning streams interact through joint attention followed by shared-stream processing, following the multimodal paradigm of FLUX and Qwen-Image \citep{flux2024,qwenimage2025}. Three-dimensional rotary positions encode spatial organization. The model predicts a latent velocity field, and the frozen VAE decoder recovers the generated volume.

\begin{wrapfigure}[31]{R}{0.49\linewidth}
\centering
\includegraphics[width=\linewidth]{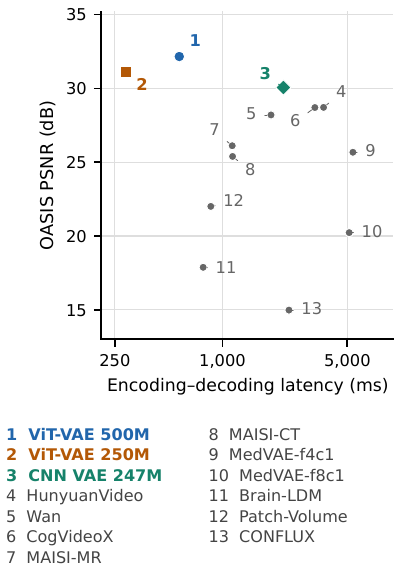}
\caption{OASIS fidelity versus encoding--decoding latency at $256^3$. Model keys are below the plot; inference policies are specified in Appendix~\ref{app:implementation}.}
\label{fig:quality-latency}
\end{wrapfigure}
The same reference pathway accommodates both synthesis and restoration. Lesion masks supply spatial guidance for generation from noise. For image-conditioned tasks, the reference is a partially observed or degraded volume: a masked image for inpainting, a blurred image for deblurring, an undersampled image for de-aliasing, or a motion-corrupted image for correction.

Because encoding is fast, these conditions are constructed in image space during training and encoded on the fly. Each presentation of a volume can receive a new mask or new degradation parameters, without caching a latent for every transformation.

\section{Experimental Setup}
Our main control is a 246.5M CNN VAE trained from scratch on the same collection, objectives and f8/C16 interface as our 249.2M and 499.3M Transformer VAEs. We use the CNN implementation of CONFLUX \citep{conflux2026}, which already has this interface; its released checkpoint is evaluated separately as CONFLUX.

\paragraph{Evaluation data and preprocessing.}
We assess brain MRI autoencoding fidelity by comparing input volumes with their decoded reconstructions on BraTS, pooled OASIS-1/OASIS-2 and IXI \citep{marcus2007oasis,marcus2010oasis,ixi_dataset}. For reconstruction evaluation, BraTS denotes 200 BraTS-GLI \citep{baid2021brats} subjects selected with a fixed random seed after excluding cases that may overlap with our training collection, with one examination and four MRI sequences per subject (800 volumes). Tumor-preservation and generation cohorts are described in Appendices~\ref{app:retention} and~\ref{app:generation}. We also report results on the internal test split, whose patients are disjoint from the training and validation splits. Cross-domain evaluations cover AMOS MRI, fastMRI knee, MSD heart and prostate MRI, and liver, lung, pancreas and kidney CT \citep{knoll2020fastmri,ji2022amos,antonelli2022msd,heller2021kits}. Volumes are reoriented, resampled to 1\,mm spacing and percentile-normalized, and reconstructions are scored on a common reference grid (Appendix~\ref{app:implementation}).

\paragraph{Metrics and coverage.}
We use PSNR as the primary reconstruction metric and report MS-SSIM, SSIM and LPIPS alongside it \citep{wang2004ssim,wang2003msssim,zhang2018lpips}, with structural metrics evaluated within a foreground bounding box. Paired confidence intervals use subject-grouped bootstrap resampling to account for multiple sequences and examinations per patient. Sequence-controlled and mask-guided synthesis are evaluated using slice-FID \citep{heusel2017fid}. Completion and restoration additionally use three-dimensional Fréchet radiomic distance (FRD) \citep{konz2026frd} within specified regions. Details are given in Appendices~\ref{app:implementation}--\ref{app:domains} for reconstruction and Appendix~\ref{app:generation} for generation.

\paragraph{Baselines and timing.}
We compare released medical autoencoders and three video VAEs at their native compression rates (Table~\ref{tab:brain}; Appendix~\ref{app:interfaces}). Training uses eight NVIDIA H200 GPUs; resource measurements use one GPU of the same type. Encoding--decoding latency uses fixed-size synthetic volumes and batch size one. All models are benchmarked using 16-bit inference (BF16 or FP16, depending on the implementation). Appendix~\ref{app:implementation} specifies timing boundaries and model-specific inference policies.

\section{Results}
\subsection{Faithful compression of medical volumes}
Strong spatial compression preserves the detail of brain MRI. The 500M Transformer VAE achieves the highest PSNR on all four brain test sets (Table~\ref{tab:brain}), reaching 32.16\,dB on OASIS compared with 28.72\,dB for the strongest released autoencoder, HunyuanVideo, which retains twice as many latent positions. Structural and perceptual metrics show consistent trends (Appendix~\ref{app:domains}), and the error maps in Figure~\ref{fig:qualitative} indicate that the improvement is distributed across tissue rather than confined to edges.

At comparable capacity, the 249.2M Transformer outperforms the 246.5M CNN on all four sets, by margins ranging from 0.09\,dB on BraTS to 1.07\,dB on OASIS, while also being substantially faster (Section~\ref{sec:context}).

We further examine whether lesion information is preserved through the bottleneck. Decoded four-sequence BraTS-GLI examinations are segmented with the frozen NS-HGlio model \citep{abayazeed2023nshglio}. Whole-tumor Dice does not decrease relative to the original images, while tumor-core and enhancing-tumor Dice decrease by 1.66 and 1.82 points, respectively, slightly less than with HunyuanVideo reconstructions (Appendix~\ref{app:retention}).

\begin{table}[!tb]
\centering\small\setlength{\tabcolsep}{4pt}\begin{tabular}{lrrrr}
\toprule
Dataset & ViT-VAE 500M & ViT-VAE 250M & CNN VAE & HunyuanVideo \\
\midrule
AMOS MRI & \textbf{34.49} & \secondbest{33.71} & 32.69 & 33.23 \\
Knee MRI & \textbf{29.20} & \secondbest{28.69} & 28.49 & 28.46 \\
Heart MRI & \textbf{34.75} & \secondbest{34.10} & 32.57 & 33.20 \\
Prostate MRI & \textbf{29.95} & \secondbest{29.05} & 27.92 & 28.52 \\
Liver CT & 35.49 & 35.22 & \secondbest{35.53} & \textbf{36.65} \\
Lung CT & \secondbest{29.09} & 28.82 & 28.82 & \textbf{29.21} \\
Pancreas CT & \secondbest{36.33} & 35.80 & 34.64 & \textbf{37.32} \\
KiTS19 CT & \secondbest{36.46} & 36.16 & 35.56 & \textbf{37.86} \\
\bottomrule
\end{tabular}

\caption{Zero-shot PSNR (dB) beyond brain MRI, without target-domain fine-tuning. CT values use volumes valid for all evaluated models; further baselines and metrics are in Appendix~\ref{app:domains}.}
\label{tab:zero-shot}
\end{table}

\subsection{Capacity, context and computational efficiency}
\label{sec:context}
\label{sec:latency}
\label{sec:train-eff}
The comparison between the two backbones also depends on input extent. When both 250M models process training-sized $128^3$ crops, their PSNR difference is negligible ($-0.005$\,dB, 95\% CI $[-0.057, 0.058]$). When both process the full volume and are scored on the same central $128^3$ region, the Transformer leads by 0.749\,dB ($[0.677, 0.838]$). Holding the scoring region fixed shows that crop-only evaluation misses an advantage observed with full-volume processing (Appendix~\ref{app:paired}).

The same efficiency makes additional capacity affordable. At $256^3$, encoding and decoding take 288.5\,ms for the 250M Transformer and 2195.7\,ms for the CNN (Table~\ref{tab:speed}). Scaling the Transformer to 500M improves PSNR on every brain test set at a latency of 574.0\,ms, still $3.8\times$ faster than the smaller CNN (Figure~\ref{fig:quality-latency}). At $384^3$, both Transformers complete whole-volume inference, whereas the CNN exceeds GPU memory. Because attention is global, the relative advantage decreases with volume size, although it persists across all tested sizes (Appendix~\ref{app:implementation}).

A similar gap appears in optimization and in condition encoding. A reconstruction-only training step at $128^3$ takes 65.9\,ms for the 250M Transformer, 129.6\,ms for the 500M model and 508.2\,ms for the CNN, with peak memory at batch size one of 13.4, 15.3 and 47.3\,GiB, respectively (Table~\ref{tab:training-efficiency}). Encoder-only latency at $256^3$ is 56.7\,ms for the 500M model, compared with 690.7\,ms for the CNN, which lowers the overhead of constructing conditions online.

\subsection{Variable-shape inference and zero-shot transfer}
The representation learned from brain MRI transfers to other anatomies. Without fine-tuning, the 500M model achieves the highest PSNR on abdominal, knee, cardiac and prostate MRI, outperforming both the CNN and HunyuanVideo (Table~\ref{tab:zero-shot}). These volumes also differ in size and aspect ratio from the training crops (Appendix~\ref{app:shape}); coordinate-based positional encoding allows them to be processed without resizing.

\begin{wrapfigure}[35]{R}{0.43\linewidth}
\centering
\includegraphics[width=\linewidth]{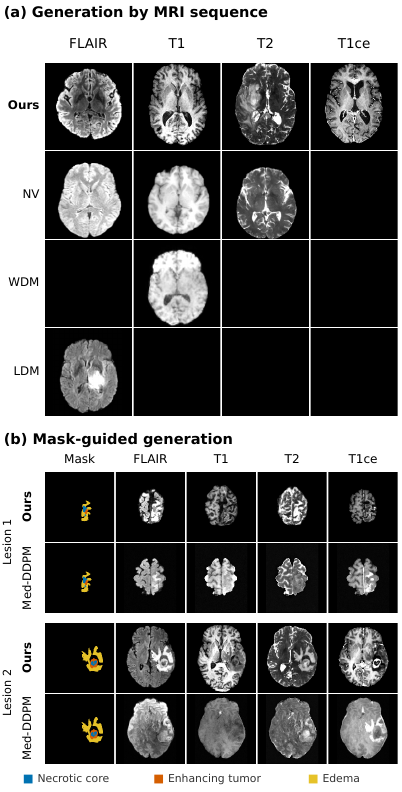}
\caption{\textbf{(a)} Sequence-controlled synthesis. \textbf{(b)} Shared-mask generation with Ours and Med-DDPM. Black cells indicate unsupported contrasts.}
\label{fig:dit-generation}
\end{wrapfigure}
Transfer to CT additionally changes the imaging modality. On the CT sets, HunyuanVideo leads by 0.1--1.4\,dB, and the 500M model ranks second on three of the four sets (Appendix~\ref{app:domains}).

\subsection{Text and spatially conditioned volumetric synthesis}
\label{sec:dit-resources}
\label{sec:generation}
The compact latent grid also reduces the sequence length processed by the generator. At a fixed latent patch size, f8/C16 makes a 2B DiT training step $1.4\times$ faster than $(4,8,8)$/C16 and $16\times$ faster than f4/C4, with approximately half the peak memory of the latter (Appendix~\ref{app:resources}). A larger DiT patch can recover the same token budget from a less compressed latent; the design question then becomes whether spatial reduction is learned by a reconstruction-trained autoencoder or by the generator's patch projection. In natural images, DC-AE shows that assigning more compression to the autoencoder improves sample quality at a fixed token count \citep{chen2024dcae}, and we adopt the same design.

Within this representation, a single 2B multimodal DiT generates complete T1, contrast-enhanced T1, T2 and FLAIR volumes. Text specifies the target sequence, and a lesion mask supplied through the reference pathway specifies tumor location and geometry (Figure~\ref{fig:dit-generation}).

Under shared lesion masks, our model achieves lower slice-FID than Med-DDPM \citep{dorjsembe2024medddpm} on all four sequences and against both reference cohorts, by approximately a factor of three on internal references (Table~\ref{tab:generation}). Med-DDPM, which additionally receives a brain-region label, achieves higher mask adherence (Dice 0.847 vs.\ 0.824), whereas our model attains higher tumor recall (Appendix~\ref{app:generation}). For sequence-controlled synthesis, our model achieves the lowest FID on internal T1 and FLAIR and is the only evaluated model that supports all four sequences, while NV-Generate \citep{nvidia2026nvbrain} performs better against BraTS references. Because the generators were trained on different data, comparisons are made within each reference cohort.

\subsection{Image-conditioned completion and restoration}
\label{sec:restoration}
\begin{figure}[!t]
\makeatletter\def\@captype{figure}\makeatother
\centering\includegraphics[width=\linewidth]{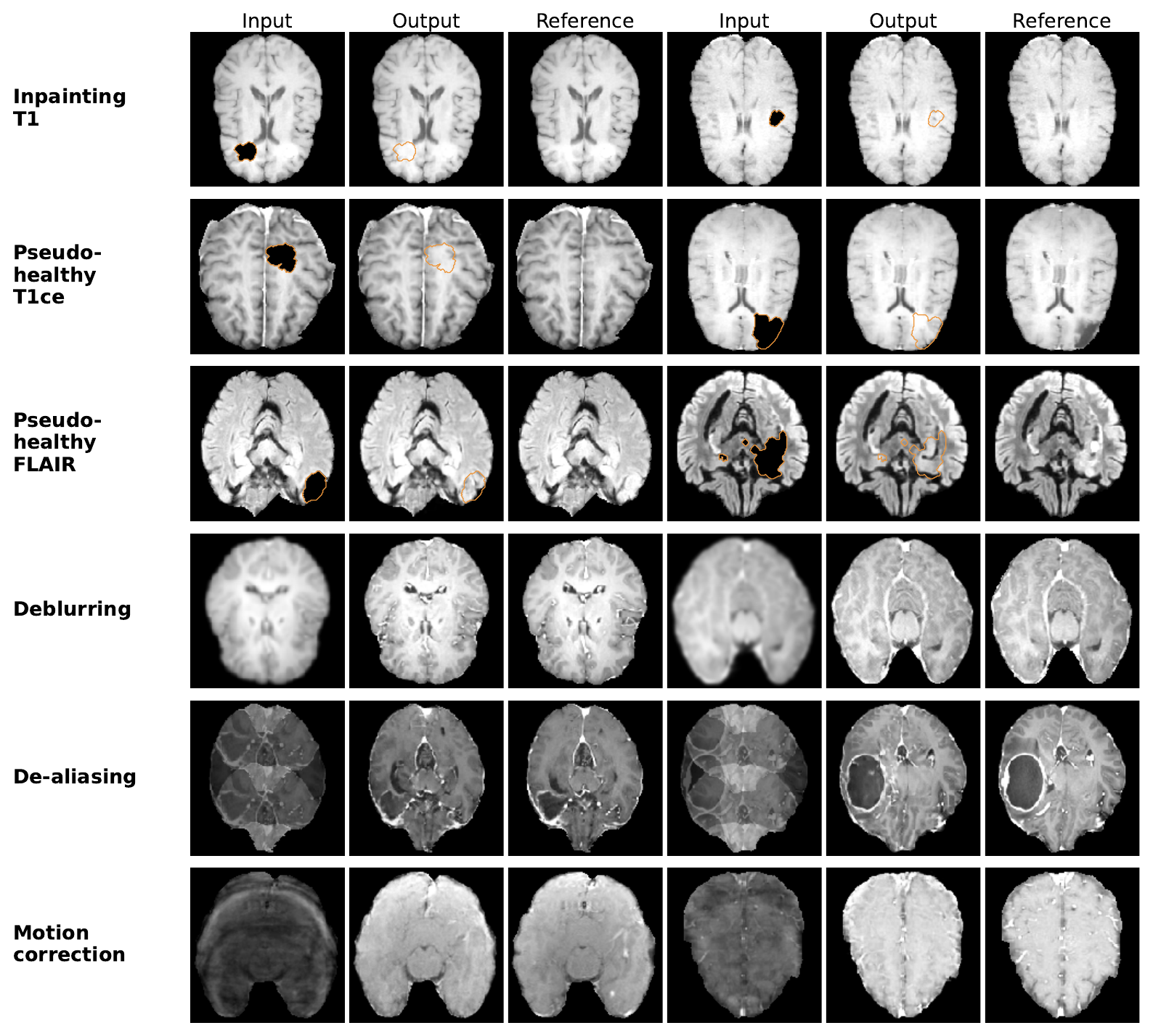}
\caption{Image-conditioned completion and restoration, with input, output and reference for two cases per row. De-aliasing displays use the reference brain-support mask; pseudo-healthy references retain the lesion.}
\label{fig:dit-image-tasks}

\end{figure}

An observed volume provides a richer condition than a sequence label or a mask. Encoded on the target grid, it specifies the existing anatomy, allowing the same reference pathway to support missing-region completion and recovery from simulated degradations (Figure~\ref{fig:dit-image-tasks}).

The model recovers a substantial part of the information lost to degradation. Motion correction improves PSNR from 10.4 to 16.1\,dB, and de-aliasing from 10.8 to 15.9\,dB. Deblurring reduces LPIPS from 0.81 to 0.22 and FID from 76 to 24 at a cost of 0.7\,dB in PSNR, reflecting a tradeoff between the reported perceptual/distributional measures and paired pixel fidelity (Appendix~\ref{app:generation}).

Inpainting exhibits a similar trade-off. We compare with Zhang25 \citep{zhang2025inpainting}, the BraTS 2025 inpainting winner \citep{kofler2023bratsinpainting}, on identical holes. After compositing both methods with the observed image outside the hole, our model has lower whole-image FID (0.18 vs.\ 0.28) and lower hole-region FRD (5.32 vs.\ 7.54), whereas Zhang25 has higher paired reconstruction fidelity (24.9 vs.\ 20.3\,dB PSNR; Appendix~\ref{app:restoration-comparison}). Raw complete-output FID favors Zhang25 (0.277 vs.\ 8.255; Table~\ref{tab:inpainting-fid-views}), indicating that the distributional comparison depends on whether changes outside the hole are included. Filling a real lesion region instead yields a healthy-appearing completion without a paired target; these pseudo-healthy examples are provided in Appendix~\ref{app:generation}.

\begin{table}[!tb]
\centering\small
\caption{Slice-FID $\downarrow$ against original internal or BraTS references. Baselines: NV-Generate, MONAI BraTS-LDM, WDM and Med-DDPM \citep{nvidia2026nvbrain,monai_brats_bundle,friedrich2024wdm,dorjsembe2024medddpm}. Dashes indicate unsupported sequences.}
\label{tab:generation}
\setlength{\tabcolsep}{4pt}\begin{tabular}{llrrrr}
\toprule
Model & Reference & T1 & T1ce & T2 & FLAIR \\
\midrule
\multicolumn{6}{l}{\textit{Modality-controlled}} \\
NV-Generate & Internal & 47.73 & -- & \textbf{44.19} & 48.20 \\
NV-Generate & BraTS & \textbf{66.80} & -- & \textbf{34.12} & \textbf{38.88} \\
MONAI BraTS-LDM & Internal & -- & -- & -- & 63.84 \\
MONAI BraTS-LDM & BraTS & -- & -- & -- & 51.66 \\
WDM & Internal & 123.36 & -- & -- & -- \\
WDM & BraTS & 102.34 & -- & -- & -- \\
Ours & Internal & \textbf{46.74} & 34.99 & 48.17 & \textbf{29.70} \\
Ours & BraTS & 77.28 & 60.85 & 54.93 & 56.47 \\
\midrule
\multicolumn{6}{l}{\textit{Mask-guided}} \\
Med-DDPM & Internal & 156.87 & 95.36 & 104.75 & 96.64 \\
Med-DDPM & BraTS & 111.83 & 91.83 & 105.19 & 80.52 \\
Ours & Internal & \textbf{45.75} & \textbf{32.58} & \textbf{42.57} & \textbf{29.15} \\
Ours & BraTS & \textbf{76.89} & \textbf{61.27} & \textbf{52.09} & \textbf{56.89} \\
\bottomrule
\end{tabular}

\end{table}

\section{Discussion and Limitations}
\label{sec:discussion}
Our results indicate a favorable reconstruction--cost tradeoff for volumetric Transformer autoencoders: at comparable size, the Transformer matches the CNN on training-sized crops, leads under full-volume input and runs at substantially lower latency, and a larger decoder improves reconstruction while retaining this advantage. Encoding efficiency also changes the role of the VAE in generative training. With a fast encoder, the VAE becomes part of the training loop rather than a preprocessing step: random holes, lesion masks and simulated degradations are generated as volumes are sampled, and text-, mask- and image-conditioned tasks share a single generator.

Unlike instruction-based image editing, the conditions in completion and restoration are partial or degraded observations of the same patient, and the generated volume is assessed by its consistency with them. The same conditioning pathway could also support studies of super-resolution, slice interpolation, missing-contrast imputation, cross-modality translation, and acquisition harmonization, although extending the checkpoint to them would require suitable conditioning data and additional training. The framework therefore provides a starting point for broader medical image processing; practical use would further require training on real degradations, consistency with the acquired measurements, uncertainty quantification and clinical validation.

Because the brain-trained model transfers to other MRI anatomies and, to a lesser extent, to CT, it may serve as an inexpensive initialization where large in-domain corpora are unavailable.

\paragraph{Limitations.} The private training data limit exact replication. The comparable-capacity comparison controls the training collection and latent interface, but does not match total training compute or isolate individual architectural components. Comparisons with released models additionally retain differences in training data and objectives. At matched token count, f8/C16 with patch size two and f4/C4 with patch size four have similar DiT training cost (Appendix~\ref{app:resources}); we do not establish superior generation quality for f8/C16 under a matched budget, and DC-AE motivates rather than demonstrates this choice in our setting. Mask adherence relies on a segmenter that expects co-registered sequences, whereas our four sequences are generated independently. Slice-FID and regional FRD measure distributions, not whole-brain anatomical consistency. CT is evaluated on normalized intensities rather than Hounsfield units, and the restoration tasks use simulated degradations.

\section{Conclusion}
\label{sec:conclusion}
We presented a Transformer VAE with eightfold compression along each spatial axis for volumetric medical imaging. It supports full-volume processing at substantially lower latency than a comparably sized CNN, and its larger variant achieves the highest brain MRI fidelity among the evaluated autoencoders while remaining several times faster. Fast encoding further allows image conditions to be constructed online, so that a single multimodal DiT supports synthesis, completion and restoration.

\label{maintextend}

\section*{Reproducibility Statement}
We describe the architecture, training settings, compression accounting, evaluation preprocessing, metrics, paired statistics and timing boundaries. Code is provided at \url{https://github.com/OncoMMAI/brainmri-vit-vae3d}. VAE weights are available through the repository. The DiT weights will not be released due to the potential for generative models to disclose sensitive training data. The private training corpus and subject-level records will not be distributed.

\section*{Ethics Statement}
The study was approved by the UCSF Institutional Review Board (protocols 22-37795 and 25-44823). The internal MRI data and subject-level records are private and will not be released. The model is intended for research; clinical use would require further validation.

\section*{AI Use Statement}
Generative AI tools assisted with literature search, experiment planning, benchmark code development, analysis of results, table preparation, and manuscript writing and editing. The authors verified all code, results and text and take full responsibility for the content. Figure~\ref{fig:architecture} was generated using an AI image-generation tool from an author-specified description of the model architecture.

\bibliography{references}
\bibliographystyle{iclr2027_conference}
\appendix

\section{Training Settings}
\label{app:training}
BrainTumor-256K contains 256,277 three-dimensional sequence entries from 8,098 patients across 71,942 examinations. Training, validation and test splits are disjoint by patient; internal reconstruction results are reported on the test split. Training uses $128^3$ crops and BF16 mixed precision. The perceptual loss applies SqueezeNet \citep{iandola2016squeezenet} LPIPS \citep{zhang2018lpips} to a random 25\% of 2D slices. The convolutional comparison uses the same training collection and latent interface; its batch size is smaller because of its higher activation memory (Table~\ref{tab:training-efficiency}).

\begin{table}[!htbp]
\centering\small\setlength{\tabcolsep}{4pt}
\caption{Training settings of the 499.3M VAE.}
\label{tab:training-settings}
\begin{tabular}{ll}
\toprule
Setting & Value \\
\midrule
Training steps & 462k \\
Examples processed & 29.6M \\
Training compute & $\approx$900--1,000 H200 GPU-hours \\
Effective batch size & 64 \\
Optimizer & AdamW \citep{loshchilov2019adamw}, $(\beta_1,\beta_2)=(0.9,0.95)$, weight decay 0.01 \\
Learning rate & $5\times10^{-5}$, constant, no warmup \\
$\lambda_{\mathrm{vox}}$ / $\lambda_{\mathrm{str}}$ / $\lambda_{\mathrm{perc}}$ & 10 / 1 / 1 \\
KL weight $\beta$ & $10^{-5}$, fixed \\
EMA decay & 0.9995 \\
Checkpoint selection & EMA validation PSNR on 128 central crops \\
\bottomrule
\end{tabular}

\end{table}

\section{Compression and DiT Token Accounting}
\label{app:compression}
\begin{table}[!htbp]
\centering\small
\caption{Analytical latent sizes for a grayscale $256^3$ input. C denotes latent channels.}
\begin{tabular}{lrrrr}
\toprule
Reduction / channels & Spatial positions & Scalar elements & Grid ratio & Element ratio\\
\midrule
$(8,8,8)$ / C16 & 32,768 & 524,288 & 512 & 32\\
$(4,8,8)$ / C16 & 65,536 & 1,048,576 & 256 & 16\\
$(8,8,8)$ / C32 & 32,768 & 1,048,576 & 512 & 16\\
$(8,8,8)$ / C1 & 32,768 & 32,768 & 512 & 512\\
\bottomrule
\end{tabular}
\end{table}
For a downstream latent patch of $p_D\times p_H\times p_W$, the token count is
\begin{equation}
N_{\mathrm{DiT}}=\frac{DHW}{f_Df_Hf_Wp_Dp_Hp_W},
\end{equation}
assuming divisible dimensions. The channel count $C$ affects only the DiT input and output projections.

\section{Evaluated Autoencoder Interfaces}
\label{app:interfaces}
\begin{table}[!htbp]
\centering\scriptsize\setlength{\tabcolsep}{3pt}
\caption{Native latent interfaces and parameter counts of the evaluated autoencoders. Ratios are relative to a single-channel volume.}
\begin{tabular}{lrllrrr}
\toprule
Model & Params M & Latent type & Factors & $C$ & Grid ratio & Element ratio \\
\midrule
ViT-VAE 500M & 499.3 & Gaussian & 8/8/8 & 16 & 512 & 32.0 \\
ViT-VAE 250M & 249.2 & Gaussian & 8/8/8 & 16 & 512 & 32.0 \\
CNN VAE 247M & 246.5 & Gaussian & 8/8/8 & 16 & 512 & 32.0 \\
HunyuanVideo & 246.5 & Gaussian & 4/8/8 & 16 & 256 & 16.0 \\
Wan & 126.9 & Gaussian & 4/8/8 & 16 & 256 & 16.0 \\
CogVideoX & 215.6 & Gaussian & 4/8/8 & 16 & 256 & 16.0 \\
MedVAE-f4c1 & 161.0 & Gaussian & 4/4/4 & 1 & 64 & 64.0 \\
MedVAE-f8c1 & 246.0 & Gaussian & 8/8/8 & 1 & 512 & 512.0 \\
Patch-Volume & 34.8 & VQ embedding & 8/8/8 & 8 & 512 & 64.0 \\
MAISI-MR & 20.9 & Gaussian & 4/4/4 & 4 & 64 & 16.0 \\
MAISI-CT & 20.9 & Gaussian & 4/4/4 & 4 & 64 & 16.0 \\
Brain-LDM & 13.8 & Gaussian & 8/8/8 & 3 & 512 & 170.7 \\
CONFLUX & 246.5 & Gaussian & 8/8/8 & 16 & 512 & 32.0 \\
\bottomrule
\end{tabular}

\end{table}
Video VAEs receive grayscale volumes replicated across their RGB input channels. For Patch-Volume, the table reports the compression of its eight-dimensional VQ embeddings.

\section{Architecture and Evaluation Details}
\label{app:implementation}
\paragraph{Evaluation preprocessing and metrics.}
The BraTS evaluation set samples 200 eligible BraTS-GLI subjects with a fixed seed, using one examination and four sequences per subject (800 volumes). OASIS combines T1-weighted inputs from OASIS-1 and OASIS-2; IXI uses a multi-sequence subset. Reference volumes use RAS orientation, 1\,mm resampling and normalization between the 0.5th and 99.5th intensity percentiles. For CT, all-model comparisons use the common valid-volume subset. The trained CNN uses tiling where whole-volume inference exceeds memory; released CONFLUX and MedVAE use tiled or sliding-window paths, and Patch-Volume uses native patch encoding.

MS-SSIM uses the three-scale MONAI implementation \citep{cardoso2022monai} with kernel seven and weights $(0.0448,0.2856,0.3001)$. Structural metrics use a foreground bounding box with four-voxel padding. Foreground extraction thresholds normalized intensity at 0.01 and applies morphological closing and hole filling. LPIPS uses a slice-based SqueezeNet implementation.

\paragraph{Baseline implementations.}
Brain-LDM uses the UK Biobank autoencoder released with MONAI's brain image synthesis bundle \citep{pinaya2022brain,monai_brain_bundle}. MAISI-MR and MAISI-CT denote the released NV-Generate-MR and NV-Generate-CT autoencoders, respectively \citep{guo2024maisi,zhao2026maisiv2,nvidia2025nvmr,nvidia_nvct}. Patch-Volume denotes the autoencoder from 3D MedDiffusion \citep{wang2024meddiffusion}. MedVAE-f4c1 and MedVAE-f8c1 are the two evaluated MedVAE variants \citep{medvae2025}. We also evaluate CONFLUX and the released CogVideoX, HunyuanVideo and Wan VAEs \citep{yang2024cogvideox,kong2024hunyuan,wan2025}.

\paragraph{Resource measurements.}
All resource measurements use one GPU and, except for the largest-batch probe, batch size one. Our models run with BF16 autocast and FP32 parameters; released models use their native 16-bit policy. Latency is measured with CUDA events over 25 calls after five warmups (five calls for released models at $384^3$); baseline adapter calls may include host transfers and padding. Training probes run twenty AdamW steps with an MSE-only objective on $128^3$ inputs (Table~\ref{tab:training-efficiency}). FLOPs count executed operators, including fused attention, with one MAC as two FLOPs.

\begin{table}[!htbp]
\centering\scriptsize\resizebox{\linewidth}{!}{\begin{tabular}{llrrrl}
\toprule
Model & Params (M) & $128^3$ ms & $256^3$ ms & $384^3$ ms & Policy \\
\midrule
ViT-VAE 500M & 499.3 & 38.7 & 574.0 & 4960.3 & Whole volume \\
ViT-VAE 250M & 249.2 & 19.6 & 288.5 & 2485.5 & Whole volume \\
CNN VAE 247M & 246.5 & 249.7 & 2195.7 & OOM & Whole volume \\
HunyuanVideo & 246.5 & 476.5 & 3685.3 & 12222.7 & Whole volume \\
Wan & 126.9 & 440.1 & 1872.2 & 5947.6 & Whole volume \\
CogVideoX & 215.6 & 447.6 & 3295.0 & 10905.5 & Whole volume \\
MAISI-MR & 20.9 & 253.4 & 1135.7 & 3429.9 & Whole volume \\
MAISI-CT & 20.9 & 260.9 & 1140.8 & 3441.1 & Whole volume \\
MedVAE-f4c1 & 161.0 & 421.4 & 5379.0 & 19412.1 & Sliding window \\
MedVAE-f8c1 & 246.0 & 316.8 & 5127.2 & 16686.3 & Sliding window \\
Brain-LDM & 13.8 & 97.1 & 781.0 & 2686.1 & Whole volume \\
Patch-Volume & 34.8 & 110.1 & 861.7 & 58378.1 & Patch encoding \\
CONFLUX & 246.5 & 265.5 & 2359.4 & 28418.1 & Whole / tiled \\
\bottomrule
\end{tabular}
}
\caption{Encoding--decoding latency (ms) of all evaluated autoencoders under their native inference policies.}
\end{table}
\begin{table}[!htbp]
\centering\small\begin{tabular}{llrrr}
\toprule
Model & Edge & Encoder ms & Peak GiB & TFLOPs \\
\midrule
ViT-VAE 500M & 128 & 4.3 & 2.12 & 0.688 \\
ViT-VAE 500M & 256 & 56.7 & 3.40 & 20.893 \\
ViT-VAE 500M & 384 & 499.1 & 6.90 & 211.526 \\
ViT-VAE 250M & 128 & 4.3 & 1.14 & 0.686 \\
ViT-VAE 250M & 256 & 57.4 & 2.43 & 20.880 \\
ViT-VAE 250M & 384 & 501.3 & 5.92 & 211.482 \\
CNN VAE 247M & 128 & 80.2 & 4.99 & 13.249 \\
CNN VAE 247M & 256 & 690.7 & 33.05 & 107.914 \\
CNN VAE 247M & 384 & 2782.8 & 109.20 & -- \\
\bottomrule
\end{tabular}

\caption{Encoder-only latency, peak allocated memory and TFLOPs.}
\end{table}
\begin{table}[!htbp]
\centering\small\begin{tabular}{llrrr}
\toprule
Model & Edge & Encode & Decode & Full \\
\midrule
ViT-VAE 500M & 128 & 0.688 & 6.151 & 6.838 \\
ViT-VAE 500M & 256 & 20.893 & 187.743 & 208.636 \\
ViT-VAE 500M & 384 & 211.526 & 1902.743 & 2114.269 \\
ViT-VAE 250M & 128 & 0.686 & 2.730 & 3.415 \\
ViT-VAE 250M & 256 & 20.880 & 83.409 & 104.289 \\
ViT-VAE 250M & 384 & 211.482 & 845.556 & 1057.038 \\
CNN VAE 247M & 128 & 13.249 & 34.707 & 47.956 \\
CNN VAE 247M & 256 & 107.914 & 279.578 & 387.493 \\
CNN VAE 247M & 384 & -- & -- & -- \\
\bottomrule
\end{tabular}

\caption{Executed-operation counts (TFLOPs) at batch one. Dashes indicate runs that exceed memory.}
\end{table}

\begin{table}[!htbp]
\centering\small\setlength{\tabcolsep}{4pt}\begin{tabular}{lrrrrr}
\toprule
Backbone & Params (M) & Step (ms) & Samples/s & B1 peak alloc. (GiB) & Largest tested batch \\
\midrule
ViT-VAE 500M & 499.3 & 129.6 & 7.7 & 15.3 & 8 \\
ViT-VAE 250M & 249.2 & 65.9 & 15.2 & 13.4 & 16 \\
CNN VAE 247M & 246.5 & 508.2 & 2.0 & 47.3 & 2 \\
\bottomrule
\end{tabular}

\caption{Training cost of the three reconstruction backbones under the shared optimization probe. Step time, throughput and peak allocated memory are measured at batch size one; the last column is the largest batch size that fits in memory.}
\label{tab:training-efficiency}
\end{table}

\section{Paired Reconstruction Statistics}
\begin{figure}[!htbp]
\centering\includegraphics[width=\linewidth]{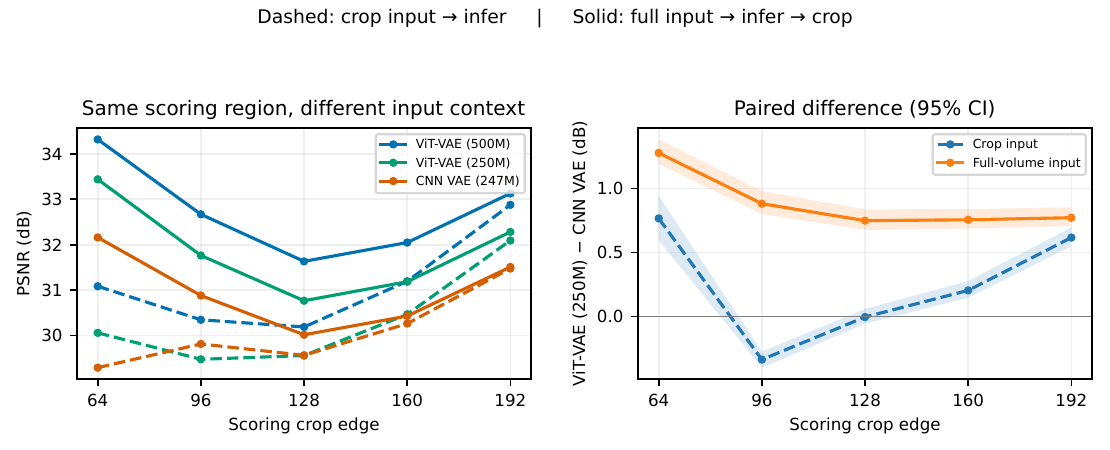}
\caption{Input context and reconstruction fidelity. Dashed: crop, then infer. Solid: infer on the full volume, then crop. Right: paired PSNR difference between the 250M ViT-VAE and the CNN VAE with 95\% confidence intervals.}
\label{fig:context}
\end{figure}

We separate the input context from the scoring region on 400 internal test volumes from 222 patients. Each model either processes a central crop directly or processes the full volume, after which its output is cropped to the same region. At $128^3$, the paired PSNR difference between the 249.2M Transformer and the 246.5M CNN is $-0.005$\,dB (95\% CI $[-0.057,0.058]$) with crop input and $+0.749$\,dB ($[0.677,0.838]$) with full-volume input; whole-volume scoring gives $+0.775$\,dB ($[0.709,0.856]$). The larger Transformer further improves fidelity across the tested contexts (Figure~\ref{fig:context}). The comparison does not identify which factor produces this difference, such as global attention, positional encoding, boundary effects or training dynamics.

\label{app:paired}
Intervals use a patient-grouped bootstrap with 2,000 replicates, resampling all volumes of a patient together.

\section{Input Shape Coverage}
\label{app:shape}
Shapes are measured after preprocessing and before model-specific padding. Axis ranges give marginal minima and maxima; aspect ratio is the longest axis divided by the shortest.
\begin{table}[!htbp]
\centering\scriptsize\setlength{\tabcolsep}{4pt}
\caption{Reference-volume shape coverage.}
\begin{tabular}{lrrrr}
\toprule
Dataset & $n$ & Distinct shapes & Axis ranges & Max. aspect \\
\midrule
BraTS & 800 & 1 & 155--155/240--240/240--240 & 1.55 \\
OASIS & 562 & 1 & 160--160/256--256/256--256 & 1.60 \\
IXI & 200 & 4 & 144--180/240--240/240--240 & 1.67 \\
BrainTumor-256K & 400 & 1 & 189--189/233--233/197--197 & 1.23 \\
AMOS MRI & 101 & 36 & 175--500/120--400/346--500 & 3.75 \\
Knee MRI & 320 & 21 & 84--150/140--140/140--140 & 1.67 \\
Heart MRI & 30 & 11 & 123--192/300--400/300--400 & 3.25 \\
Prostate MRI & 96 & 9 & 44--86/192--232/192--232 & 5.27 \\
Liver CT & 200 & 196 & 74--1800/237--512/285--512 & 6.92 \\
Lung CT & 95 & 95 & 169--636/306--500/306--500 & 2.37 \\
Pancreas CT & 200 & 196 & 120--670/277--500/277--500 & 3.22 \\
KiTS19 CT & 200 & 200 & 145--755/224--533/224--533 & 3.45 \\
\bottomrule
\end{tabular}

\end{table}

\section{Multi-metric Autoencoding}
\label{app:domains}
The tables below report the three medically trained models. The CT tables additionally include released baselines on volumes valid for all models (131 liver, 92 lung, 195 pancreas and 153 KiTS19), with CNN tiling where needed.
\begin{table}[!htbp]
\centering\small
\caption{PSNR (dB) across imaging domains. Bold and boxed values mark the best and second-best of the three models.}
\begin{tabular}{lrrr}
\toprule
Dataset & ViT-VAE 500M & ViT-VAE 250M & CNN VAE 247M \\
\midrule
BraTS & \textbf{33.53} & \secondbest{32.55} & 32.46 \\
OASIS & \textbf{32.16} & \secondbest{31.13} & 30.06 \\
IXI & \textbf{35.46} & \secondbest{34.08} & 33.82 \\
BrainTumor-256K & \textbf{33.80} & \secondbest{32.95} & 32.18 \\
AMOS MRI & \textbf{34.49} & \secondbest{33.71} & 32.69 \\
Knee MRI & \textbf{29.20} & \secondbest{28.69} & 28.49 \\
Heart MRI & \textbf{34.75} & \secondbest{34.10} & 32.57 \\
Prostate MRI & \textbf{29.95} & \secondbest{29.05} & 27.92 \\
Liver CT & \secondbest{35.49} & 35.22 & \textbf{35.53} \\
Lung CT & \textbf{29.09} & \secondbest{28.82} & \secondbest{28.82} \\
Pancreas CT & \textbf{36.33} & \secondbest{35.80} & 34.64 \\
KiTS19 CT & \textbf{36.46} & \secondbest{36.16} & 35.56 \\
\bottomrule
\end{tabular}
\end{table}
\begin{table}[!htbp]
\centering\small
\caption{MS-SSIM across imaging domains. Bold and boxed values mark the best and second-best of the three models.}
\begin{tabular}{lrrr}
\toprule
Dataset & ViT-VAE 500M & ViT-VAE 250M & CNN VAE 247M \\
\midrule
BraTS & \textbf{0.9968} & \secondbest{0.9962} & 0.9942 \\
OASIS & \textbf{0.9972} & \secondbest{0.9965} & 0.9949 \\
IXI & \textbf{0.9985} & \secondbest{0.9980} & 0.9971 \\
BrainTumor-256K & \textbf{0.9960} & \secondbest{0.9953} & 0.9940 \\
AMOS MRI & \textbf{0.9939} & \secondbest{0.9931} & 0.9920 \\
Knee MRI & \textbf{0.9764} & \secondbest{0.9741} & 0.9730 \\
Heart MRI & \textbf{0.9924} & \secondbest{0.9918} & 0.9895 \\
Prostate MRI & \textbf{0.9849} & \secondbest{0.9834} & 0.9809 \\
Liver CT & \secondbest{0.9885} & \secondbest{0.9885} & \textbf{0.9897} \\
Lung CT & \textbf{0.9671} & 0.9660 & \secondbest{0.9668} \\
Pancreas CT & \textbf{0.9919} & \textbf{0.9919} & \secondbest{0.9911} \\
KiTS19 CT & 0.9899 & \secondbest{0.9902} & \textbf{0.9908} \\
\bottomrule
\end{tabular}
\end{table}
\begin{table}[!htbp]
\centering\small
\caption{SSIM across imaging domains. Bold and boxed values mark the best and second-best of the three models.}
\begin{tabular}{lrrr}
\toprule
Dataset & ViT-VAE 500M & ViT-VAE 250M & CNN VAE 247M \\
\midrule
BraTS & \textbf{0.9684} & \secondbest{0.9636} & 0.9543 \\
OASIS & \textbf{0.9694} & \secondbest{0.9627} & 0.9529 \\
IXI & \textbf{0.9853} & \secondbest{0.9810} & 0.9760 \\
BrainTumor-256K & \textbf{0.9666} & \secondbest{0.9621} & 0.9553 \\
AMOS MRI & \textbf{0.9543} & \secondbest{0.9480} & 0.9425 \\
Knee MRI & \textbf{0.8035} & \secondbest{0.7918} & 0.7890 \\
Heart MRI & \textbf{0.9468} & \secondbest{0.9418} & 0.9305 \\
Prostate MRI & \textbf{0.8775} & \secondbest{0.8687} & 0.8595 \\
Liver CT & \secondbest{0.9141} & \secondbest{0.9141} & \textbf{0.9207} \\
Lung CT & \secondbest{0.7611} & 0.7582 & \textbf{0.7637} \\
Pancreas CT & \textbf{0.9400} & \secondbest{0.9394} & 0.9349 \\
KiTS19 CT & 0.9282 & \secondbest{0.9293} & \textbf{0.9346} \\
\bottomrule
\end{tabular}
\end{table}
\begin{table}[!htbp]
\centering\small
\caption{LPIPS across imaging domains. Bold and boxed values mark the best and second-best of the three models.}
\begin{tabular}{lrrr}
\toprule
Dataset & ViT-VAE 500M & ViT-VAE 250M & CNN VAE 247M \\
\midrule
BraTS & \textbf{0.0399} & 0.0474 & \secondbest{0.0460} \\
OASIS & \textbf{0.0485} & \secondbest{0.0575} & 0.0636 \\
IXI & \textbf{0.0296} & 0.0371 & \secondbest{0.0356} \\
BrainTumor-256K & \textbf{0.0635} & 0.0726 & \secondbest{0.0708} \\
AMOS MRI & \textbf{0.0872} & \secondbest{0.0957} & 0.1033 \\
Knee MRI & \secondbest{0.2005} & 0.2067 & \textbf{0.1930} \\
Heart MRI & \textbf{0.1051} & \secondbest{0.1101} & 0.1172 \\
Prostate MRI & \textbf{0.1228} & \secondbest{0.1402} & 0.1643 \\
Liver CT & \secondbest{0.1385} & 0.1442 & \textbf{0.1132} \\
Lung CT & 0.1930 & \secondbest{0.1873} & \textbf{0.1739} \\
Pancreas CT & \secondbest{0.1154} & 0.1214 & \textbf{0.1087} \\
KiTS19 CT & \secondbest{0.1367} & 0.1401 & \textbf{0.1029} \\
\bottomrule
\end{tabular}
\end{table}

\begin{table}[!htbp]
\centering\small
\caption{PSNR (dB) on common valid CT volumes, using each implementation's feasible inference policy.}
\begin{tabular}{lrrrr}
\toprule
Model & Liver & Lung & Pancreas & KiTS19 \\
\midrule
ViT-VAE 500M & 35.49 & \secondbest{29.09} & \secondbest{36.33} & \secondbest{36.46} \\
ViT-VAE 250M & 35.22 & 28.82 & 35.80 & 36.16 \\
CNN VAE 247M & 35.53 & 28.82 & 34.64 & 35.56 \\
HunyuanVideo & \textbf{36.65} & \textbf{29.21} & \textbf{37.32} & \textbf{37.86} \\
Wan & \secondbest{35.54} & 28.60 & 36.30 & 36.33 \\
CONFLUX & 28.42 & 25.95 & 26.42 & 25.30 \\
MedVAE-f4c1 & 30.54 & 27.20 & 30.26 & 29.73 \\
MedVAE-f8c1 & 20.65 & 17.99 & 19.96 & 19.75 \\
Patch-Volume & 29.14 & 24.75 & 28.16 & 28.38 \\
MAISI-MR & 27.56 & 24.72 & 22.76 & 28.25 \\
MAISI-CT & 27.42 & 23.04 & 23.40 & 28.18 \\
Brain-LDM & 7.98 & 9.64 & 9.16 & 7.21 \\
\bottomrule
\end{tabular}

\end{table}
\begin{table}[!htbp]
\centering\small
\caption{MS-SSIM on common valid CT volumes, using each implementation's feasible inference policy.}
\begin{tabular}{lrrrr}
\toprule
Model & Liver & Lung & Pancreas & KiTS19 \\
\midrule
ViT-VAE 500M & 0.9885 & \secondbest{0.9671} & \secondbest{0.9919} & 0.9899 \\
ViT-VAE 250M & 0.9885 & 0.9660 & \secondbest{0.9919} & 0.9902 \\
CNN VAE 247M & \secondbest{0.9897} & 0.9668 & 0.9911 & 0.9908 \\
HunyuanVideo & \textbf{0.9908} & \textbf{0.9688} & \textbf{0.9933} & \textbf{0.9924} \\
Wan & 0.9889 & 0.9642 & 0.9915 & \secondbest{0.9909} \\
CONFLUX & 0.9376 & 0.9089 & 0.9185 & 0.9360 \\
MedVAE-f4c1 & 0.9755 & 0.9569 & 0.9828 & 0.9779 \\
MedVAE-f8c1 & 0.8360 & 0.7633 & 0.8355 & 0.8281 \\
Patch-Volume & 0.9547 & 0.9150 & 0.9573 & 0.9508 \\
MAISI-MR & 0.9604 & 0.9229 & 0.9352 & 0.9683 \\
MAISI-CT & 0.9597 & 0.9155 & 0.9370 & 0.9675 \\
Brain-LDM & 0.3824 & 0.3815 & 0.4577 & 0.4063 \\
\bottomrule
\end{tabular}

\end{table}
\begin{table}[!htbp]
\centering\small
\caption{SSIM on common valid CT volumes, using each implementation's feasible inference policy.}
\begin{tabular}{lrrrr}
\toprule
Model & Liver & Lung & Pancreas & KiTS19 \\
\midrule
ViT-VAE 500M & 0.9141 & \secondbest{0.7611} & \secondbest{0.9400} & 0.9282 \\
ViT-VAE 250M & 0.9141 & 0.7582 & 0.9394 & 0.9293 \\
CNN VAE 247M & \secondbest{0.9207} & \textbf{0.7637} & 0.9349 & \secondbest{0.9346} \\
HunyuanVideo & \textbf{0.9244} & \textbf{0.7637} & \textbf{0.9466} & \textbf{0.9396} \\
Wan & 0.9124 & 0.7402 & 0.9338 & 0.9278 \\
CONFLUX & 0.7393 & 0.5504 & 0.6779 & 0.7675 \\
MedVAE-f4c1 & 0.8714 & 0.7118 & 0.9007 & 0.8884 \\
MedVAE-f8c1 & 0.6868 & 0.4922 & 0.6895 & 0.6936 \\
Patch-Volume & 0.8046 & 0.6139 & 0.8301 & 0.8113 \\
MAISI-MR & 0.7957 & 0.5682 & 0.7381 & 0.8245 \\
MAISI-CT & 0.7982 & 0.5674 & 0.7417 & 0.8279 \\
Brain-LDM & 0.1611 & 0.1357 & 0.2196 & 0.1699 \\
\bottomrule
\end{tabular}

\end{table}
\begin{table}[!htbp]
\centering\small
\caption{LPIPS on common valid CT volumes, using each implementation's feasible inference policy.}
\begin{tabular}{lrrrr}
\toprule
Model & Liver & Lung & Pancreas & KiTS19 \\
\midrule
ViT-VAE 500M & 0.1385 & 0.1930 & 0.1154 & 0.1367 \\
ViT-VAE 250M & 0.1442 & 0.1873 & 0.1214 & 0.1401 \\
CNN VAE 247M & \textbf{0.1132} & 0.1739 & 0.1087 & \secondbest{0.1029} \\
HunyuanVideo & \secondbest{0.1150} & \secondbest{0.1713} & \textbf{0.0851} & \textbf{0.1018} \\
Wan & 0.1221 & \textbf{0.1662} & \secondbest{0.0936} & 0.1128 \\
CONFLUX & 0.4377 & 0.6297 & 0.4362 & 0.4752 \\
MedVAE-f4c1 & 0.1773 & 0.1973 & 0.1989 & 0.2079 \\
MedVAE-f8c1 & 0.4498 & 0.5393 & 0.4746 & 0.5136 \\
Patch-Volume & 0.2193 & 0.2389 & 0.2633 & 0.2852 \\
MAISI-MR & 0.1894 & 0.2366 & 0.2823 & 0.1862 \\
MAISI-CT & 0.2102 & 0.2983 & 0.2889 & 0.2043 \\
Brain-LDM & 1.4832 & 1.2653 & 1.3933 & 1.5630 \\
\bottomrule
\end{tabular}

\end{table}

\section{Tumor-region Preservation Protocol and Results}
\begin{table}[!htbp]
\centering\small
\caption{Tumor segmentation Dice (\%) on original and reconstructed BraTS-GLI inputs. WT, TC and ET denote whole tumor, tumor core and enhancing tumor.}
\label{tab:retention}
\begin{tabular}{lrrr}
\toprule
Input to segmenter & WT & TC & ET \\
\midrule
Original & 83.26 & 84.92 & 82.98 \\
ViT-VAE 500M & 84.22 & \textbf{83.27} & \textbf{81.16} \\
HunyuanVideo & \textbf{84.52} & 82.88 & 80.68 \\
\bottomrule
\end{tabular}

\end{table}
\label{app:retention}
\paragraph{Cohort and comparison.}
We use 165 preoperative patients from a fixed BraTS-GLI validation partition of the BraTS benchmark \citep{menze2015brats}, with one preoperative examination per patient. Each examination includes non-contrast T1, contrast-enhanced T1, T2 and FLAIR images with a tumor annotation. The 500M VAE and the released HunyuanVideo VAE reconstruct each modality independently using posterior means; the frozen segmenter then receives the four aligned reconstructed modalities.

\paragraph{Shared preprocessing.}
Inputs are registered and skull-stripped on a 1\,mm grid and N4-corrected \citep{tustison2010n4}. Each modality is clipped to its 0.5th--99.5th percentiles and scaled to $[0,1]$; the same constants restore intensities for every input condition. NS-HGlio's own preprocessing is disabled.

\paragraph{Region definitions and scoring.}
Reference labels 1, 2 and 3 represent non-enhancing tumor core, surrounding non-enhancing FLAIR hyperintensity, and enhancing tumor; predicted labels 1, 2 and 3 represent necrosis, enhancing tumor and edema. Reference WT/TC/ET sets are therefore $\{1,2,3\}/\{1,3\}/\{3\}$, while predicted sets are $\{1,2,3\}/\{1,2\}/\{2\}$. Dice is computed per region; an empty prediction for a present region scores zero, and absent reference regions are excluded. ET is absent in seven patients.

\begin{table}[!htbp]
\centering\small
\caption{Patient-paired Dice differences (percentage points, ours minus comparator) with 95\% patient-bootstrap intervals.}
\label{tab:retention-paired}
\begin{tabular}{llrrl}
\toprule
Comparator & Region & Patients & Difference & 95\% interval \\
\midrule
Original & WT & 165 & +0.96 & [+0.12, +1.76] \\
Original & TC & 165 & -1.66 & [-2.87, -0.64] \\
Original & ET & 158 & -1.82 & [-2.84, -0.96] \\
HunyuanVideo & WT & 165 & -0.30 & [-0.78, +0.32] \\
HunyuanVideo & TC & 165 & +0.38 & [+0.10, +0.69] \\
HunyuanVideo & ET & 158 & +0.48 & [+0.02, +0.91] \\
\bottomrule
\end{tabular}

\end{table}

\section{DiT Resource Benchmark Protocol}
\label{app:resources}
We profile complete DiT training steps (forward, backward and AdamW) on synthetic latents for a $192\times192\times144$ image, using one GPU, batch size one, BF16 autocast and fused scaled dot-product attention \citep{dao2022flash}, with five warmup and twenty measured steps; VAE execution and data loading are excluded. Within each DiT size (633M, 988M and 2.01B), interfaces differ only in the channel-dependent input and output projections.

\begin{table}[!htbp]
\caption{DiT training cost by latent interface. p is the DiT patch size (two unless stated); SD is the within-run standard deviation of step time.}
\label{tab:dit-resources}
\centering\scriptsize\setlength{\tabcolsep}{4pt}\begin{tabular}{llrrrrr}
\toprule
DiT & Interface & Tokens & Step ms & SD ms & Alloc. GiB & Res. GiB \\
\midrule
633M & f8/C16 (ours, CNN VAE) & 1,296 & 64.8 & 0.3 & 9.85 & 10.93 \\
633M & (4,8,8)/C16 (video VAEs) & 2,592 & 70.9 & 5.8 & 11.31 & 12.40 \\
633M & f4/C4 (MAISI), p2 & 10,368 & 946.2 & 79.0 & 19.70 & 20.66 \\
988M & f8/C16 (ours, CNN VAE) & 1,296 & 67.4 & 0.2 & 14.85 & 18.12 \\
988M & (4,8,8)/C16 (video VAEs) & 2,592 & 78.9 & 0.1 & 16.55 & 18.99 \\
2.01B & f8/C16 (ours, CNN VAE) & 1,296 & 113.2 & 0.2 & 30.23 & 36.26 \\
2.01B & (4,8,8)/C16 (video VAEs) & 2,592 & 161.6 & 15.6 & 34.30 & 38.98 \\
2.01B & f4/C4 (MAISI), p2 & 10,368 & 1828.5 & 343.1 & 57.92 & 60.74 \\
2.01B & f4/C4 (MAISI), p4 & 1,296 & 118.3 & 26.4 & 30.24 & 36.25 \\
\bottomrule
\end{tabular}

\end{table}

\section{Conditional Generation Protocol}
\label{app:generation}
All tasks use one 2B multimodal DiT of width 1728 with the frozen 499.3M VAE; a $192\times192\times144$ image maps to a $16\times24\times24\times18$ latent.

Modality-controlled generation disables all reference-image inputs and conditions on the requested sequence. Mask-guided generation adds a tumor-label map through the reference-conditioning pathway. Sampling uses 50 Euler steps with a power time schedule of exponent three. Classifier-free text guidance \citep{ho2022cfg} is five for modality-controlled synthesis; mask-guided synthesis uses text and reference guidance of four and six. Guidance settings and generator weights were selected on validation data before evaluation.

\begin{table}[!htbp]
\centering\small
\caption{Tasks supported by the shared generator. Each task also specifies the target MRI sequence.}
\label{tab:task-interface}
\begin{tabular}{lll}
\toprule
Task & Spatial condition & Synthesis target\\
\midrule
Sequence control & None & Requested MRI contrast\\
Lesion-mask guidance & Tumor-label map & MRI following lesion geometry\\
Inpainting & Observed image and hole & Completed image\\
Deblurring & Blurred image & Sharp image\\
De-aliasing & Undersampled-image reference & Corrected image\\
Motion correction & Motion-corrupted image & Corrected image\\
\bottomrule
\end{tabular}
\end{table}

\paragraph{Completion and restoration evaluation.}
We evaluate the same 2B generator on 100 held-out subjects, with one examination per subject. Healthy-region inpainting uses T1-weighted MRI and a shared intracranial hole for both methods; the comparison uses Zhang25. Deblurring, de-aliasing and motion correction use contrast-enhanced T1-weighted MRI with simulated degradations. All tasks use 50 Euler steps, fixed seeds, text guidance four and reference guidance one. Deblurring uses Gaussian $\sigma=1.5$; de-aliasing uses factor-three k-space undersampling; motion correction uses simulated periodic head motion. PSNR is computed within the indicated evaluation mask; structural and perceptual measures use its bounding-box context. Confidence intervals resample subjects; undefined metrics are excluded per metric. For pseudo-healthy synthesis, PSNR outside the lesion relative to the original image is 25.78\,dB.

\begin{table}[!tbp]
\centering\small
\caption{Image-conditioned task comparisons. Inpainting uses the reference outside the hole for both methods; restoration compares raw outputs with degraded inputs. FID uses whole-volume slices; FRD ($\log d^2$) uses hole/brain regions, respectively.}
\label{tab:restoration}
\setlength{\tabcolsep}{4pt}\begin{tabular}{llrrrr}
\toprule
Task & Method & LPIPS $\downarrow$ & MS-SSIM $\uparrow$ & FID $\downarrow$ & FRD $\downarrow$ \\
\midrule
Inpainting & Zhang25 & 0.069 & \textbf{0.961} & 0.28 & 7.54 \\
& Ours & \textbf{0.063} & 0.941 & \textbf{0.18} & \textbf{5.32} \\
Deblurring & Degraded input & 0.808 & 0.959 & 76.20 & 25.77 \\
& Ours & \textbf{0.222} & \textbf{0.965} & \textbf{23.95} & \textbf{5.83} \\
De-aliasing & Degraded input & 0.847 & 0.627 & 198.30 & 8.66 \\
& Ours & \textbf{0.670} & \textbf{0.759} & \textbf{154.29} & \textbf{6.52} \\
Motion correction & Degraded input & 0.303 & 0.803 & 62.10 & 7.63 \\
& Ours & \textbf{0.192} & \textbf{0.899} & \textbf{21.85} & \textbf{6.25} \\
\bottomrule
\end{tabular}

\end{table}

\begin{table}[!htbp]
\centering\small\begin{tabular}{llrl}
\toprule
Task & Metric & Mean & 95\% interval \\
\midrule
Inpainting & PSNR & 20.262 & [19.660, 20.863] \\
Inpainting & MS-SSIM & 0.918 & [0.906, 0.930] \\
Inpainting & SSIM & 0.763 & [0.741, 0.784] \\
Inpainting & LPIPS & 0.175 & [0.165, 0.187] \\
Deblurring & PSNR & 20.303 & [20.097, 20.513] \\
Deblurring & MS-SSIM & 0.965 & [0.962, 0.967] \\
Deblurring & SSIM & 0.818 & [0.809, 0.827] \\
Deblurring & LPIPS & 0.222 & [0.206, 0.238] \\
De-aliasing & PSNR & 15.859 & [15.451, 16.252] \\
De-aliasing & MS-SSIM & 0.759 & [0.751, 0.767] \\
De-aliasing & SSIM & 0.335 & [0.325, 0.344] \\
De-aliasing & LPIPS & 0.670 & [0.649, 0.692] \\
Motion correction & PSNR & 16.100 & [15.721, 16.481] \\
Motion correction & MS-SSIM & 0.899 & [0.896, 0.903] \\
Motion correction & SSIM & 0.747 & [0.740, 0.754] \\
Motion correction & LPIPS & 0.192 & [0.187, 0.198] \\
\bottomrule
\end{tabular}

\caption{Subject-bootstrap intervals for paired completion and restoration metrics.}
\end{table}

\paragraph{Distributional task evaluation.}
FID pools Inception-v3 \citep{szegedy2016inception} features from full-volume axial slices between 20\% and 85\% of depth, using per-slice min--max normalization. FRD uses the official implementation \citep{konz2026frd} with 1,014 extracted radiomic features, standardized against the real cohort; features constant in that cohort are removed. We report $\log(d^2)$, where $d^2$ is the squared Fr\'{e}chet distance. Radiomic features are extracted within the shared hole for inpainting and the brain region for restoration. For inpainting in the main table, both methods are composited with the reference outside the hole before FID, LPIPS and MS-SSIM evaluation; hole FRD uses the generated region. Restoration uses raw outputs.

\paragraph{Paired task comparisons.}
\label{app:restoration-comparison}
Table~\ref{tab:restoration-comparison} compares restoration with its degraded input and inpainting with Zhang25 on identical targets and holes. With the reference restored outside each hole, LPIPS is 0.063 for Ours and 0.069 for Zhang25 (paired 95\% interval $[-0.015,0.002]$).
\begin{table}[!htbp]
\centering\small
\begin{tabular}{llrrrr}
\toprule
Task & Method & PSNR $\uparrow$ & MS-SSIM $\uparrow$ & SSIM $\uparrow$ & LPIPS $\downarrow$ \\
\midrule
Inpainting & Zhang25 & \textbf{24.87} & \textbf{0.960} & \textbf{0.929} & \textbf{0.075} \\
Inpainting & Ours & 20.26 & 0.918 & 0.763 & 0.175 \\
Deblurring & Degraded input & \textbf{20.97} & 0.959 & 0.795 & 0.808 \\
Deblurring & Ours & 20.30 & \textbf{0.965} & \textbf{0.818} & \textbf{0.222} \\
De-aliasing & Degraded input & 10.77 & 0.627 & \textbf{0.442} & 0.847 \\
De-aliasing & Ours & \textbf{15.86} & \textbf{0.759} & 0.335 & \textbf{0.670} \\
Motion correction & Degraded input & 10.43 & 0.803 & 0.717 & 0.303 \\
Motion correction & Ours & \textbf{16.10} & \textbf{0.899} & \textbf{0.747} & \textbf{0.192} \\
\bottomrule
\end{tabular}

\caption{Paired completion and restoration comparisons. Inpainting uses the missing region; the remaining tasks use the brain region. Raw model outputs are evaluated for both methods.}
\label{tab:restoration-comparison}
\end{table}

\paragraph{Inpainting context and de-aliasing post-processing.}
Zhang25 preserves voxels outside the hole: on the common evaluation grid, its outside-hole mean absolute error relative to the reference is 0.0013, with a mean correlation of 0.9965. The DiT, in contrast, decodes the complete volume; Table~\ref{tab:preservation-frd} reports its outside-region FRD. The reference-outside composite applies the same observed context to both outputs and isolates the synthesized region, whereas raw whole-volume FID additionally reflects reconstruction outside the hole (Table~\ref{tab:inpainting-fid-views}).
\begin{table}[!htbp]
\centering\small\begin{tabular}{lrr}
\toprule
Inpainting evaluation & Ours FID & Zhang25 FID \\
\midrule
Raw complete output & 8.2548 & 0.2770 \\
Reference-outside composite & 0.1776 & 0.2768 \\
\bottomrule
\end{tabular}

\caption{Inpainting FID under raw-output and shared reference-outside evaluation. Both methods use the same subjects and holes.}
\label{tab:inpainting-fid-views}
\end{table}
De-aliasing displays multiply input and output by the reference brain support ($x>0.01$). Table~\ref{tab:dealias-postprocess} reports this view separately: brain-mask PSNR is unchanged, while structural metrics and FID improve.
\begin{table}[!htbp]
\centering\small\begin{tabular}{lrrrrr}
\toprule
De-aliasing output & PSNR & MS-SSIM & SSIM & LPIPS & FID \\
\midrule
Raw & 15.859 & 0.759 & 0.335 & 0.670 & 154.29 \\
Brain-support masked & 15.859 & 0.891 & 0.726 & 0.226 & 21.23 \\
\bottomrule
\end{tabular}

\caption{De-aliasing with and without reference brain-support masking. Raw outputs are the primary evaluation.}
\label{tab:dealias-postprocess}
\end{table}

\paragraph{Lesion-region restoration.}
Table~\ref{tab:lesion-restoration} reports paired metrics in lesion regions; small regions can make multiscale or perceptual scores undefined.
\begin{table}[!htbp]
\centering\scriptsize\setlength{\tabcolsep}{4pt}\begin{tabular}{llrrrrr}
\toprule
Task & Method & PSNR & MS-SSIM & SSIM & LPIPS & Valid $n$ (P/M/S/L) \\
\midrule
Deblurring & Input & \textbf{25.785} & 0.935 & \textbf{0.705} & 0.953 & 100/84/99/93 \\
Deblurring & Ours & 23.615 & \textbf{0.939} & 0.682 & \textbf{0.349} & 100/84/99/93 \\
De-aliasing & Input & 12.378 & 0.620 & \textbf{0.454} & 0.730 & 100/84/99/93 \\
De-aliasing & Ours & \textbf{15.737} & \textbf{0.723} & 0.404 & \textbf{0.529} & 100/84/99/93 \\
Motion correction & Input & 11.331 & 0.700 & 0.483 & 0.665 & 100/84/99/93 \\
Motion correction & Ours & \textbf{17.233} & \textbf{0.788} & \textbf{0.505} & \textbf{0.361} & 100/84/99/93 \\
\bottomrule
\end{tabular}

\caption{Lesion-region restoration fidelity. P/M/S/L give valid case counts for PSNR, MS-SSIM, SSIM and LPIPS.}
\label{tab:lesion-restoration}
\end{table}

\begin{figure}[!htbp]
\centering\includegraphics[width=\linewidth,height=0.83\textheight,keepaspectratio,trim=0 0 0 19,clip]{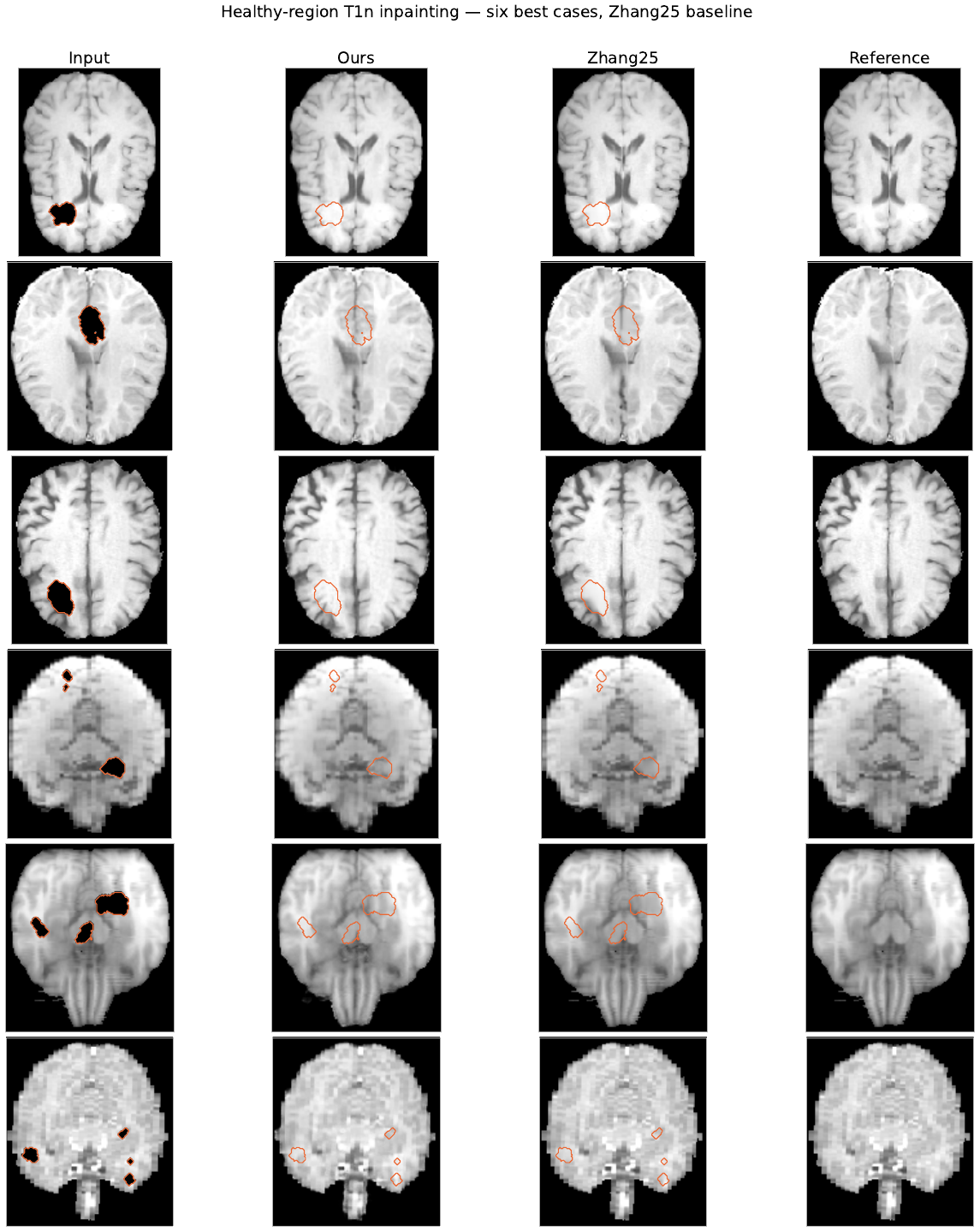}
\caption{Healthy-region T1 inpainting. Columns show input, Ours, Zhang25 and the original reference under the same missing-region condition.}
\label{fig:healthy-inpainting-baseline}
\end{figure}

\begin{figure}[!htbp]
\centering\includegraphics[width=\linewidth,height=0.87\textheight,keepaspectratio,trim=0 0 0 30,clip]{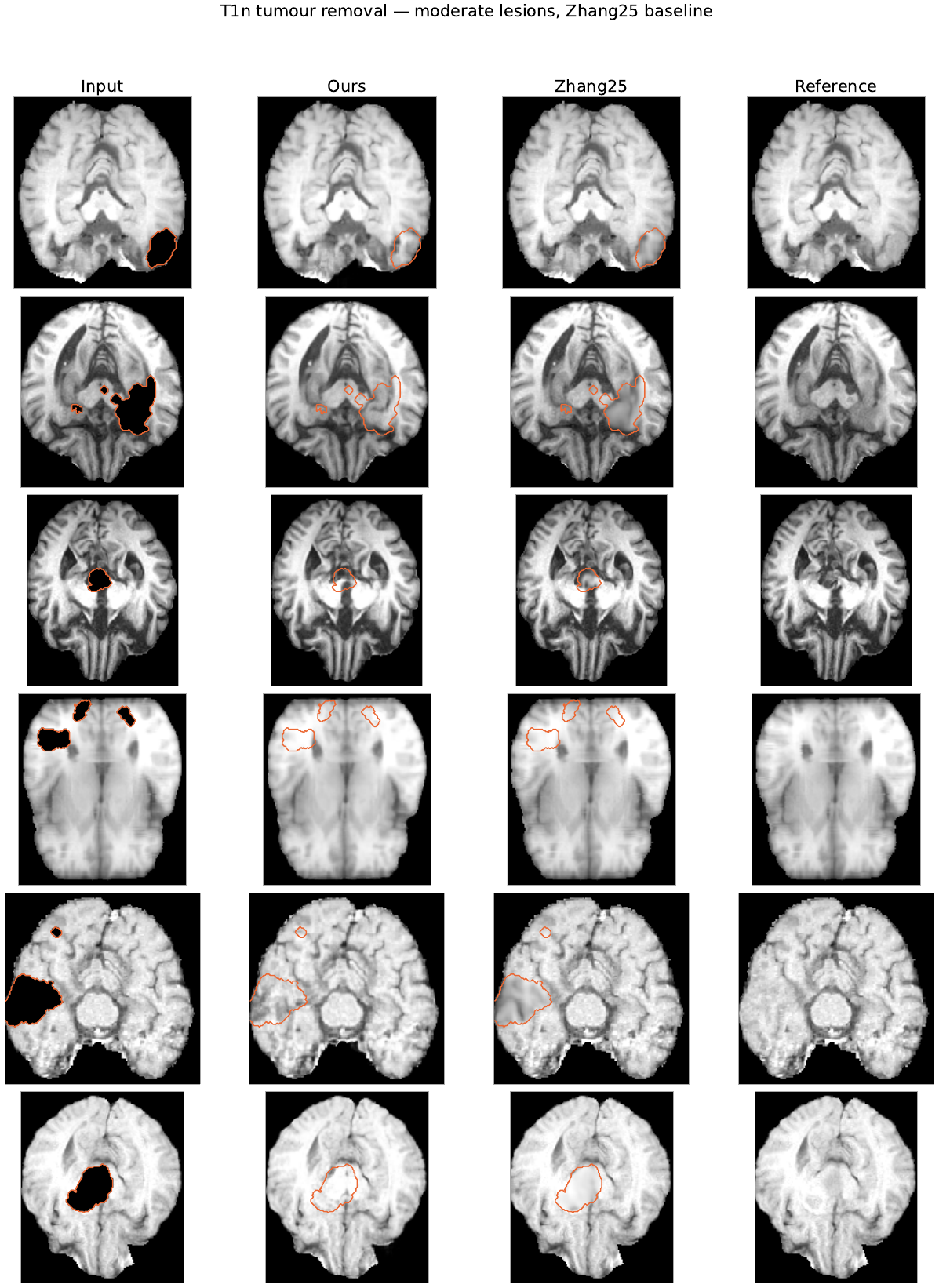}
\caption{T1 lesion-region synthesis. Columns show input, Ours, Zhang25 and the original tumor-bearing MRI.}
\label{fig:pseudohealthy-t1}
\end{figure}

\begin{figure}[!htbp]
\centering\includegraphics[width=\linewidth,height=0.87\textheight,keepaspectratio]{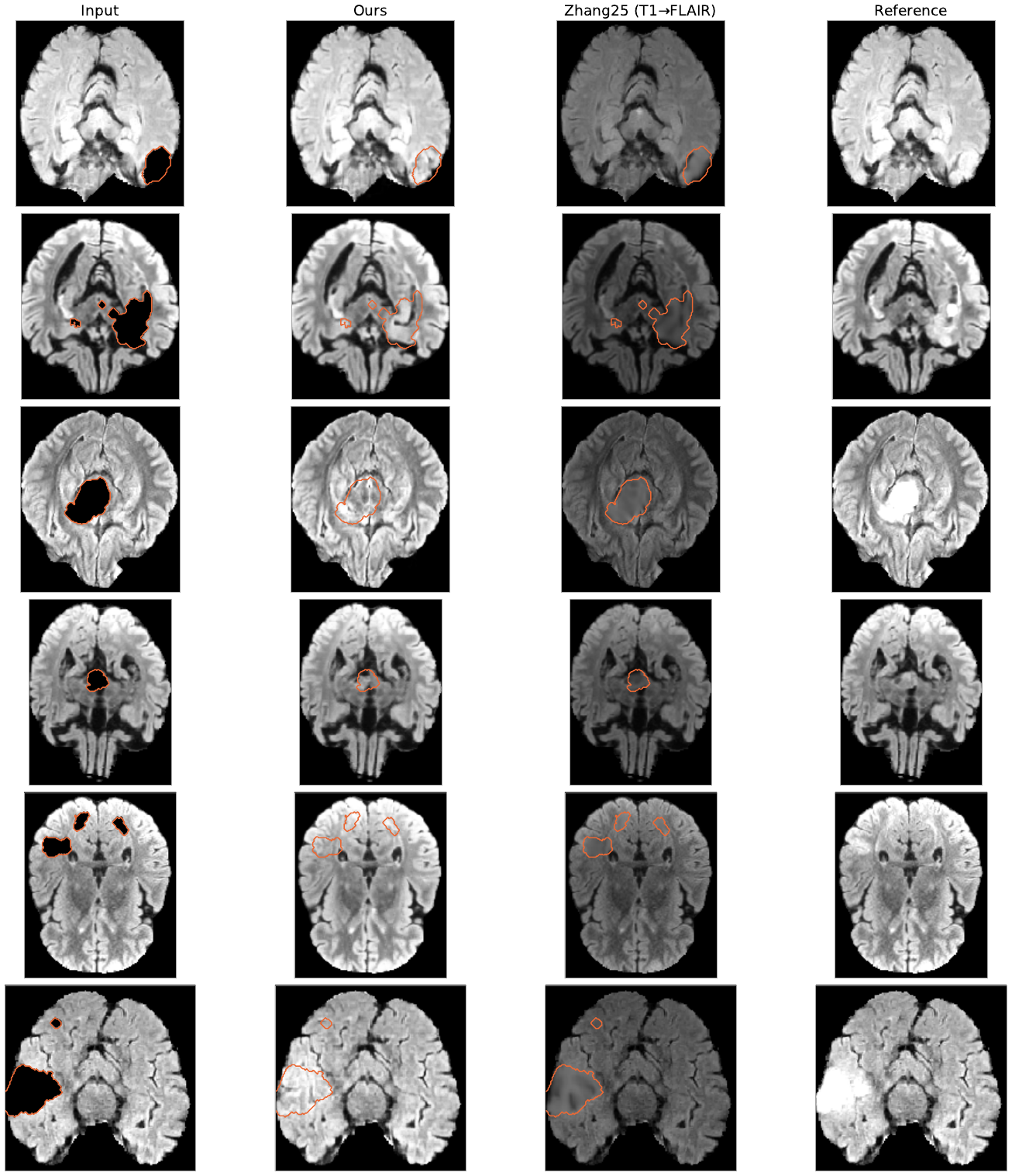}
\caption{FLAIR pseudo-healthy synthesis. Columns show input, Ours, Zhang25 (T1-trained, out of domain) and the original MRI.}
\label{fig:pseudohealthy-flair}
\end{figure}

\paragraph{Pseudo-healthy synthesis and contextual preservation.}
Lesion-conditioned completion has no paired healthy target. Table~\ref{tab:preservation-frd} reports FRD outside the generated region, as a check that full-volume decoding preserves the observed tissue. In a separate four-sequence assessment of 20 subjects, a frozen NS-HGlio segmenter \citep{abayazeed2023nshglio} is applied to original and generated images on the same grid. The residual fraction divides segmentation-detected whole-tumor volume in the generated region by that detected on the original images. Its mean is 0.209 and median 0.024 (Table~\ref{tab:pseudohealthy-residual}). Figures~\ref{fig:pseudohealthy-t1} and~\ref{fig:pseudohealthy-flair} show lesion-region examples; Zhang25 is trained on T1, so its FLAIR results are out of domain.
\begin{table}[!htbp]
\centering\small\begin{tabular}{lr}
\toprule
Task / region & FRD ($\log d^2$) $\downarrow$ \\
\midrule
Healthy-region inpainting / outside hole & 4.071 \\
Pseudo-healthy synthesis / outside lesion & 4.200 \\
\bottomrule
\end{tabular}

\caption{Outside-region FRD of our full-volume outputs on the common evaluation grid.}
\label{tab:preservation-frd}
\end{table}
\begin{table}[!htbp]
\centering\small\begin{tabular}{lrr}
\toprule
Segmentation-detected residual & Mean & Median \\
\midrule
Fraction within the conditioned region & 0.209 & 0.024 \\
Fraction over the full volume & 0.255 & 0.082 \\
\midrule
\multicolumn{3}{l}{Residual within region $\leq1\%$: 9/20; $\leq5\%$: 11/20.} \\
\bottomrule
\end{tabular}

\caption{Segmentation-detected residual tumor fractions after pseudo-healthy synthesis (20 subjects).}
\label{tab:pseudohealthy-residual}
\end{table}


\paragraph{Original-image references.}
We evaluate the released generators and both tasks of our model against internal and BraTS cohorts of original MRI, using 400 generated and 400 reference volumes per supported sequence for each reported score. The internal cohort comprises four-sequence examinations drawn from the test split, excluding examinations used for mask conditioning; the external cohort comprises BraTS subjects. Generated volumes and BraTS references are reoriented, resampled to 1\,mm spacing, percentile-normalized and center-cropped or padded to $192\times192\times144$; internal references use the preprocessed 1\,mm images on the same grid. Neither reference cohort is passed through a VAE. Slice-FID follows the protocol above \citep{heusel2017fid}; because the two cohorts differ in acquisition and preprocessing, scores are compared only within each cohort.

\paragraph{Generation tasks and conditions.}
NV-Generate-MR-Brain \citep{nvidia2026nvbrain}, MONAI BraTS-LDM \citep{monai_brats_bundle} and WDM \citep{friedrich2024wdm} use their released sampling pipelines for the supported sequences. MONAI BraTS-LDM denotes the BraTS FLAIR bundle; it is distinct from the UK Biobank Brain-LDM autoencoder used in reconstruction comparisons. Our mask-guided model and Med-DDPM \citep{dorjsembe2024medddpm} use the same 400 test-split tumor masks, each supplying the condition for all four sequences. Our model generates the four sequences independently, conditioned on the tumor mask alone. Med-DDPM additionally receives the non-tumor brain-region label required by its conditioning interface. Masks are transformed into each model's input coordinates.

\paragraph{Tumor-mask adherence.}
We apply the frozen NS-HGlio segmenter \citep{abayazeed2023nshglio} to the four generated sequences from each of the 400 shared conditioning masks. The predicted whole-tumor region is compared with the conditioning tumor region after a fixed coordinate transformation; the non-tumor brain label is excluded. Table~\ref{tab:mask-adherence} reports case-averaged overlap, precision, recall and the predicted-to-conditioning tumor-volume ratio. Our model has higher recall but over-segments (volume ratio 1.28), while Med-DDPM has higher Dice and precision. Because our four sequences are generated independently, their non-lesional anatomy can differ, which the segmenter may penalize.

\begin{table}[ht]
\centering\small
\caption{Tumor-mask adherence measured by a frozen segmenter on generated MRI. Volume ratio is predicted tumor volume divided by conditioning tumor volume; the ideal value is one.}
\label{tab:mask-adherence}
\begin{tabular}{lrrrr}
\toprule
Model & Dice $\uparrow$ & Precision $\uparrow$ & Recall $\uparrow$ & Volume ratio \\
\midrule
Ours & 0.824 & 0.766 & \textbf{0.911} & 1.275 \\
Med-DDPM & \textbf{0.847} & \textbf{0.844} & 0.867 & 1.052 \\
\bottomrule
\end{tabular}

\end{table}

\paragraph{Qualitative examples.}
In Figure~\ref{fig:dit-generation} and Appendix~\ref{app:mask-examples}, both mask-guided models use the same mask and slice across all four sequences. Outputs are resampled to a common axial grid with a 192\,mm field of view and volume-wise percentile normalization. Mask-guided slices show the largest tumor cross-section; sequence-controlled slices show the middle of the brain's superior--inferior extent.

\section{Additional Mask-guided Generation Examples}
\label{app:mask-examples}
\begin{figure}[!ht]
\centering
\includegraphics[width=\linewidth,height=0.84\textheight,keepaspectratio]{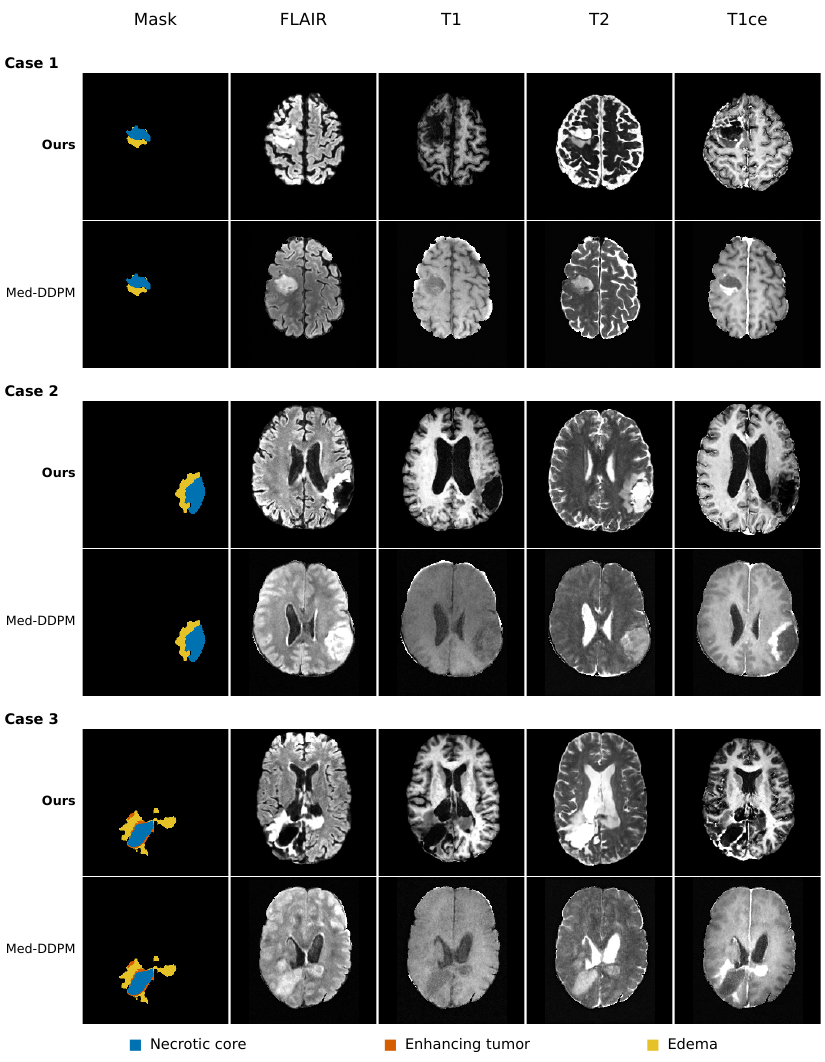}
\caption{Mask-guided generation under shared tumor conditions, cases 1--3. Each pair shows Ours above Med-DDPM across four MRI sequences.}
\label{fig:mask-comparison-one}
\end{figure}

\begin{figure}[!ht]
\centering
\includegraphics[width=\linewidth,height=0.84\textheight,keepaspectratio]{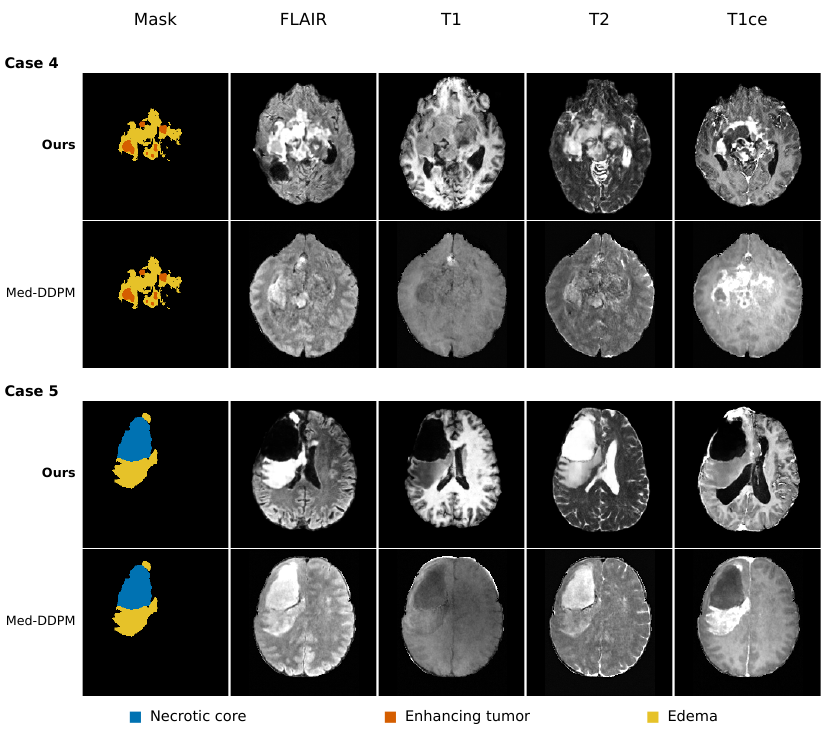}
\caption{Mask-guided generation under shared tumor conditions, cases 4--5. Layout and display settings match Figure~\ref{fig:mask-comparison-one}.}
\label{fig:mask-comparison-two}
\end{figure}

Figures~\ref{fig:mask-comparison-one} and~\ref{fig:mask-comparison-two} compare five cases spanning different conditioning tumor volumes. 


\section{Additional Reconstruction Examples}
\begin{figure}[!htbp]
\centering\includegraphics[width=\linewidth,height=0.83\textheight,keepaspectratio]{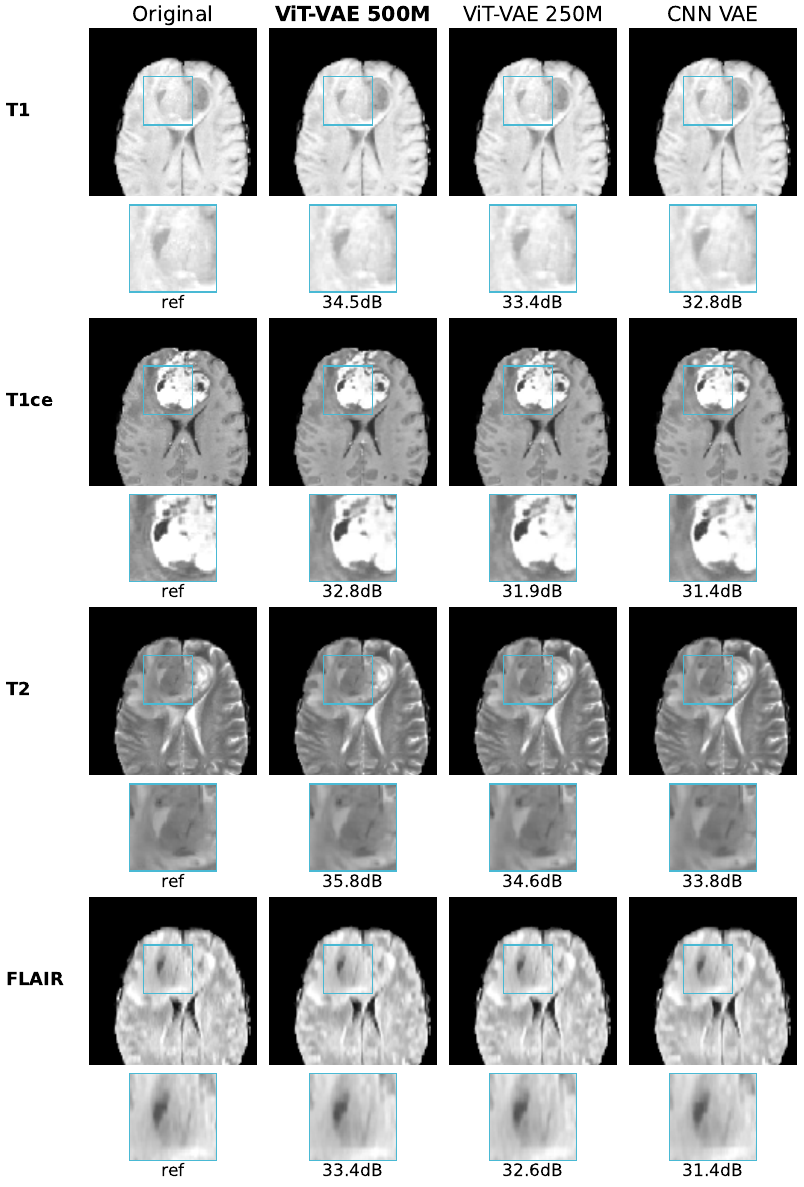}
\caption{Axial reconstructions with local detail for the two ViT-VAEs and the CNN VAE. Annotations report volume-level PSNR.}
\label{fig:vae-additional}
\end{figure}

\begin{figure}[!htbp]
\centering
\includegraphics[width=\linewidth,height=0.88\textheight,keepaspectratio]{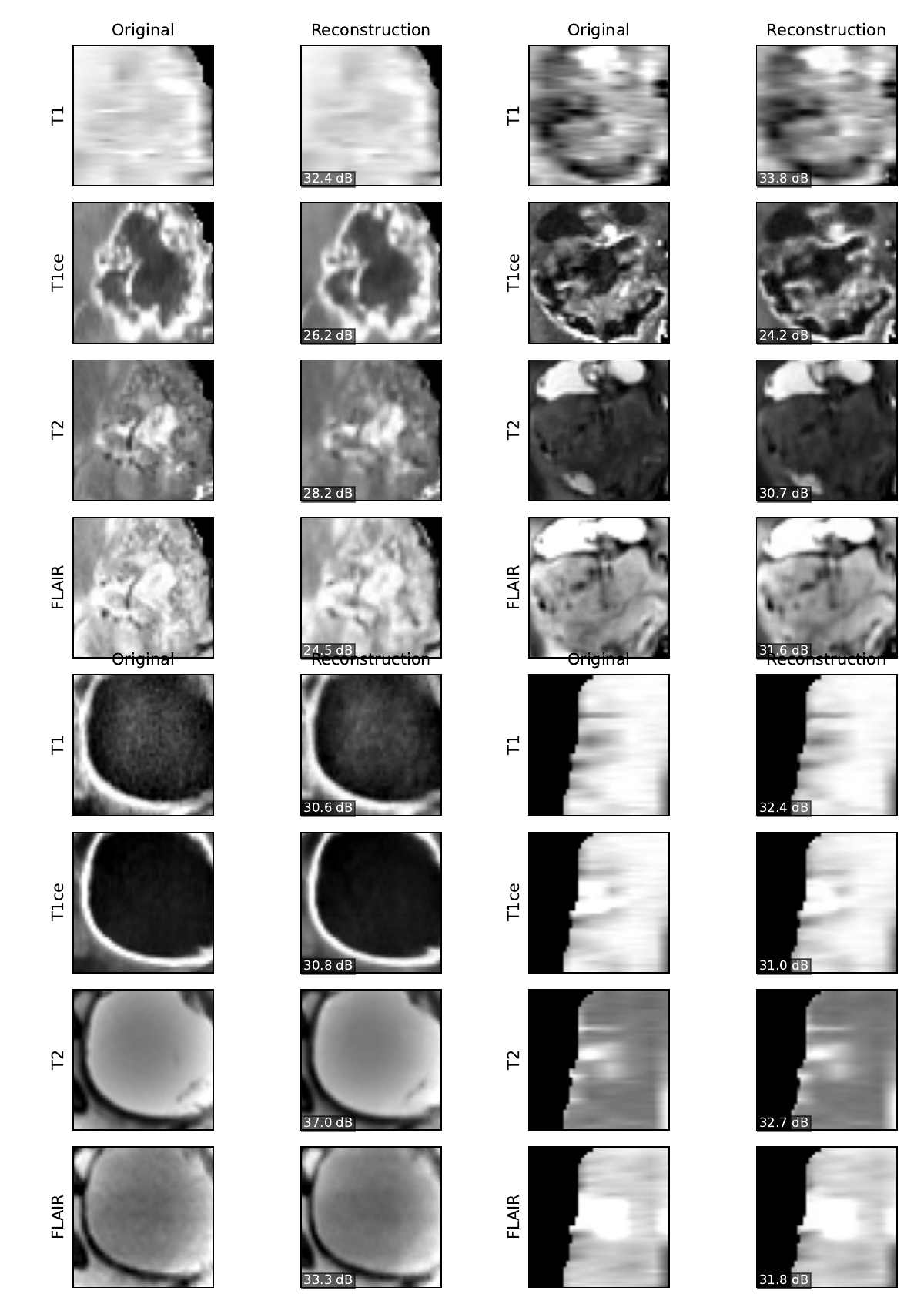}
\caption{500M original--reconstruction pairs across regional MS-SSIM quantiles. Annotations report PSNR in tumor-centered $48^3$ regions.}
\label{fig:ours-range}
\end{figure}

\label{app:examples}
Figure~\ref{fig:ours-range} shows 500M reconstructions of tumor-centered regions across four MRI sequences. Cases are drawn across regional MS-SSIM quantiles rather than selected as best examples.

\end{document}